\documentclass[longauth]{aa_gaia} % for the long lists of affiliations 
\usepackage{graphicx, array}
\usepackage[varg]{txfonts}
\usepackage{natbib,twoopt}
\usepackage{xcolor}
\usepackage[breaklinks=true]{hyperref} %% to avoid \citeads line fills
\bibpunct{(}{)}{;}{a}{}{,}             %% natbib format for A&A and ApJ
\makeatletter
  \newcommandtwoopt{\citeads}[3][][]{\href{http://adsabs.harvard.edu/abs/#3}%
    {\def\hyper@linkstart##1##2{}%
     \let\hyper@linkend\@empty\citealp[#1][#2]{#3}}}
  \newcommandtwoopt{\citepads}[3][][]{\href{http://adsabs.harvard.edu/abs/#3}%
    {\def\hyper@linkstart##1##2{}%
     \let\hyper@linkend\@empty\citep[#1][#2]{#3}}}
  \newcommandtwoopt{\citetads}[3][][]{\href{http://adsabs.harvard.edu/abs/#3}%
    {\def\hyper@linkstart##1##2{}%
     \let\hyper@linkend\@empty\citet[#1][#2]{#3}}}
  \newcommandtwoopt{\citeyearads}[3][][]%
    {\href{http://adsabs.harvard.edu/abs/#3}
    {\def\hyper@linkstart##1##2{}%
     \let\hyper@linkend\@empty\citeyear[#1][#2]{#3}}}
\makeatother
\hypersetup{colorlinks=true,linkcolor=blue,citecolor=blue,urlcolor=blue}

\newcommand\gaia{\textit{Gaia}}
\newcommand\gdr[1]{\gaia~DR#1}
\newcommand\hip{\textsc{Hipparcos}}
\newcommand\tyc{\textit{Tycho}}
\newcommand\tyctwo{\textit{Tycho}-2}
\newcommand\gaiacrftwo{\gaia-CRF2}

\newcommand\secref[1]{Sect.~\ref{#1}}
\newcommand\secsref[1]{Sects.~\ref{#1}}
\newcommand\secrefalt[1]{Section~\ref{#1}}
\newcommand\figref[1]{Fig.~\ref{#1}}
\newcommand\figsref[1]{Figs.~\ref{#1}}
\newcommand\figrefalt[1]{Figure~\ref{#1}}

\newcommand\tabref[1]{Table~\ref{#1}}

\newcommand\gdrtwototnum{\ensuremath{1\,692\,919\,135}}
\newcommand\gdrtwovarnum{\ensuremath{550\,737}}
\newcommand\gdrtwovartypednum{\ensuremath{363\,969}}
\newcommand\gdrtwovarsostotnum{\ensuremath{390\,529}}

\newcommand\gdrtwossonum{\ensuremath{14\,099}}
\newcommand\gdrtwovradnum{\ensuremath{7\,224\,631}}
\newcommand\gdrtwofivepnum{\ensuremath{1\,331\,909\,727}}
\newcommand\gdrtwotwopnum{\ensuremath{361\,009\,408}}
\newcommand\gdrtwogbandnum{\ensuremath{1\,692\,919\,135}}
\newcommand\gdrtwobpbandnum{\ensuremath{1\,381\,964\,755}}
\newcommand\gdrtworpbandnum{\ensuremath{1\,383\,551\,713}}

\newcommand\gdrtwoteffnum{\ensuremath{161\,497\,595}}
\newcommand\gdrtwoagnum{\ensuremath{87\,733\,672}}
\newcommand\gdrtwoebpminrpnum{\ensuremath{87\,733\,672}}
\newcommand\gdrtworadiusnum{\ensuremath{76\,956\,778}}
\newcommand\gdrtwolumnum{\ensuremath{76\,956\,778}}

\newcommand\gdrtwoicrfthree{\ensuremath{2820}}
\newcommand\gdrtwogaiacrftwo{\ensuremath{556\,869}}

\newcommand\masyr{\ensuremath{\text{mas~yr}^{-1}}}

\newcommand\kms{\ensuremath{\text{km~s}^{-1}}}
\newcommand\ms{\ensuremath{\text{m~s}^{-1}}}
\newcommand\gbp{\ensuremath{G_\mathrm{BP}}}
\newcommand\grp{\ensuremath{G_\mathrm{RP}}}
\newcommand\grvs{\ensuremath{G_\mathrm{RVS}}}
\newcommand\teff{\ensuremath{T_\mathrm{eff}}}
\newcommand\logg{\ensuremath{\log g}}
\newcommand\feh{\ensuremath{[\text{Fe}/\text{H}]}}
\newcommand\ag{\ensuremath{A_G}}
\newcommand\ebpminrp{\ensuremath{E(\gbp-\grp)}}
\newcommand\vrad{\ensuremath{v_\mathrm{rad}}}

\begin{document} 

\title{{\gaia} Data Release 2}

\subtitle{Summary of the contents and survey properties}

\author{
{\it Gaia} Collaboration
\and A.G.A.    ~Brown                         \inst{\ref{inst:0001}}
\and A.        ~Vallenari                     \inst{\ref{inst:0002}}
\and T.        ~Prusti                        \inst{\ref{inst:0003}}
\and J.H.J.    ~de Bruijne                    \inst{\ref{inst:0003}}
\and C.        ~Babusiaux                     \inst{\ref{inst:0005},\ref{inst:0006}}
\and C.A.L.    ~Bailer-Jones                  \inst{\ref{inst:0007}}
\and M.        ~Biermann                      \inst{\ref{inst:0008}}
\and D.W.      ~Evans                         \inst{\ref{inst:0009}}
\and L.        ~Eyer                          \inst{\ref{inst:0010}}
\and F.        ~Jansen                        \inst{\ref{inst:0011}}
\and C.        ~Jordi                         \inst{\ref{inst:0012}}
\and S.A.      ~Klioner                       \inst{\ref{inst:0013}}
\and U.        ~Lammers                       \inst{\ref{inst:0014}}
\and L.        ~Lindegren                     \inst{\ref{inst:0015}}
\and X.        ~Luri                          \inst{\ref{inst:0012}}
\and F.        ~Mignard                       \inst{\ref{inst:0017}}
\and C.        ~Panem                         \inst{\ref{inst:0018}}
\and D.        ~Pourbaix                      \inst{\ref{inst:0019},\ref{inst:0020}}
\and S.        ~Randich                       \inst{\ref{inst:0021}}
\and P.        ~Sartoretti                    \inst{\ref{inst:0005}}
\and H.I.      ~Siddiqui                      \inst{\ref{inst:0023}}
\and C.        ~Soubiran                      \inst{\ref{inst:0024}}
\and F.        ~van Leeuwen                   \inst{\ref{inst:0009}}
\and N.A.      ~Walton                        \inst{\ref{inst:0009}}
\and F.        ~Arenou                        \inst{\ref{inst:0005}}
\and U.        ~Bastian                       \inst{\ref{inst:0008}}
\and M.        ~Cropper                       \inst{\ref{inst:0029}}
\and R.        ~Drimmel                       \inst{\ref{inst:0030}}
\and D.        ~Katz                          \inst{\ref{inst:0005}}
\and M.G.      ~Lattanzi                      \inst{\ref{inst:0030}}
\and J.        ~Bakker                        \inst{\ref{inst:0014}}
\and C.        ~Cacciari                      \inst{\ref{inst:0034}}
\and J.        ~Casta\~{n}eda                 \inst{\ref{inst:0012}}
\and L.        ~Chaoul                        \inst{\ref{inst:0018}}
\and N.        ~Cheek                         \inst{\ref{inst:0037}}
\and F.        ~De Angeli                     \inst{\ref{inst:0009}}
\and C.        ~Fabricius                     \inst{\ref{inst:0012}}
\and R.        ~Guerra                        \inst{\ref{inst:0014}}
\and B.        ~Holl                          \inst{\ref{inst:0010}}
\and E.        ~Masana                        \inst{\ref{inst:0012}}
\and R.        ~Messineo                      \inst{\ref{inst:0043}}
\and N.        ~Mowlavi                       \inst{\ref{inst:0010}}
\and K.        ~Nienartowicz                  \inst{\ref{inst:0045}}
\and P.        ~Panuzzo                       \inst{\ref{inst:0005}}
\and J.        ~Portell                       \inst{\ref{inst:0012}}
\and M.        ~Riello                        \inst{\ref{inst:0009}}
\and G.M.      ~Seabroke                      \inst{\ref{inst:0029}}
\and P.        ~Tanga                         \inst{\ref{inst:0017}}
\and F.        ~Th\'{e}venin                  \inst{\ref{inst:0017}}
\and G.        ~Gracia-Abril                  \inst{\ref{inst:0052},\ref{inst:0008}}
\and G.        ~Comoretto                     \inst{\ref{inst:0023}}
\and M.        ~Garcia-Reinaldos              \inst{\ref{inst:0014}}
\and D.        ~Teyssier                      \inst{\ref{inst:0023}}
\and M.        ~Altmann                       \inst{\ref{inst:0008},\ref{inst:0058}}
\and R.        ~Andrae                        \inst{\ref{inst:0007}}
\and M.        ~Audard                        \inst{\ref{inst:0010}}
\and I.        ~Bellas-Velidis                \inst{\ref{inst:0061}}
\and K.        ~Benson                        \inst{\ref{inst:0029}}
\and J.        ~Berthier                      \inst{\ref{inst:0063}}
\and R.        ~Blomme                        \inst{\ref{inst:0064}}
\and P.        ~Burgess                       \inst{\ref{inst:0009}}
\and G.        ~Busso                         \inst{\ref{inst:0009}}
\and B.        ~Carry                         \inst{\ref{inst:0017},\ref{inst:0063}}
\and A.        ~Cellino                       \inst{\ref{inst:0030}}
\and G.        ~Clementini                    \inst{\ref{inst:0034}}
\and M.        ~Clotet                        \inst{\ref{inst:0012}}
\and O.        ~Creevey                       \inst{\ref{inst:0017}}
\and M.        ~Davidson                      \inst{\ref{inst:0073}}
\and J.        ~De Ridder                     \inst{\ref{inst:0074}}
\and L.        ~Delchambre                    \inst{\ref{inst:0075}}
\and A.        ~Dell'Oro                      \inst{\ref{inst:0021}}
\and C.        ~Ducourant                     \inst{\ref{inst:0024}}
\and J.        ~Fern\'{a}ndez-Hern\'{a}ndez   \inst{\ref{inst:0078}}
\and M.        ~Fouesneau                     \inst{\ref{inst:0007}}
\and Y.        ~Fr\'{e}mat                    \inst{\ref{inst:0064}}
\and L.        ~Galluccio                     \inst{\ref{inst:0017}}
\and M.        ~Garc\'{i}a-Torres             \inst{\ref{inst:0082}}
\and J.        ~Gonz\'{a}lez-N\'{u}\~{n}ez    \inst{\ref{inst:0037},\ref{inst:0084}}
\and J.J.      ~Gonz\'{a}lez-Vidal            \inst{\ref{inst:0012}}
\and E.        ~Gosset                        \inst{\ref{inst:0075},\ref{inst:0020}}
\and L.P.      ~Guy                           \inst{\ref{inst:0045},\ref{inst:0089}}
\and J.-L.     ~Halbwachs                     \inst{\ref{inst:0090}}
\and N.C.      ~Hambly                        \inst{\ref{inst:0073}}
\and D.L.      ~Harrison                      \inst{\ref{inst:0009},\ref{inst:0093}}
\and J.        ~Hern\'{a}ndez                 \inst{\ref{inst:0014}}
\and D.        ~Hestroffer                    \inst{\ref{inst:0063}}
\and S.T.      ~Hodgkin                       \inst{\ref{inst:0009}}
\and A.        ~Hutton                        \inst{\ref{inst:0097}}
\and G.        ~Jasniewicz                    \inst{\ref{inst:0098}}
\and A.        ~Jean-Antoine-Piccolo          \inst{\ref{inst:0018}}
\and S.        ~Jordan                        \inst{\ref{inst:0008}}
\and A.J.      ~Korn                          \inst{\ref{inst:0101}}
\and A.        ~Krone-Martins                 \inst{\ref{inst:0102}}
\and A.C.      ~Lanzafame                     \inst{\ref{inst:0103},\ref{inst:0104}}
\and T.        ~Lebzelter                     \inst{\ref{inst:0105}}
\and W.        ~L\"{ o}ffler                  \inst{\ref{inst:0008}}
\and M.        ~Manteiga                      \inst{\ref{inst:0107},\ref{inst:0108}}
\and P.M.      ~Marrese                       \inst{\ref{inst:0109},\ref{inst:0110}}
\and J.M.      ~Mart\'{i}n-Fleitas            \inst{\ref{inst:0097}}
\and A.        ~Moitinho                      \inst{\ref{inst:0102}}
\and A.        ~Mora                          \inst{\ref{inst:0097}}
\and K.        ~Muinonen                      \inst{\ref{inst:0114},\ref{inst:0115}}
\and J.        ~Osinde                        \inst{\ref{inst:0116}}
\and E.        ~Pancino                       \inst{\ref{inst:0021},\ref{inst:0110}}
\and T.        ~Pauwels                       \inst{\ref{inst:0064}}
\and J.-M.     ~Petit                         \inst{\ref{inst:0120}}
\and A.        ~Recio-Blanco                  \inst{\ref{inst:0017}}
\and P.J.      ~Richards                      \inst{\ref{inst:0122}}
\and L.        ~Rimoldini                     \inst{\ref{inst:0045}}
\and A.C.      ~Robin                         \inst{\ref{inst:0120}}
\and L.M.      ~Sarro                         \inst{\ref{inst:0125}}
\and C.        ~Siopis                        \inst{\ref{inst:0019}}
\and M.        ~Smith                         \inst{\ref{inst:0029}}
\and A.        ~Sozzetti                      \inst{\ref{inst:0030}}
\and M.        ~S\"{ u}veges                  \inst{\ref{inst:0007}}
\and J.        ~Torra                         \inst{\ref{inst:0012}}
\and W.        ~van Reeven                    \inst{\ref{inst:0097}}
\and U.        ~Abbas                         \inst{\ref{inst:0030}}
\and A.        ~Abreu Aramburu                \inst{\ref{inst:0133}}
\and S.        ~Accart                        \inst{\ref{inst:0134}}
\and C.        ~Aerts                         \inst{\ref{inst:0074},\ref{inst:0136}}
\and G.        ~Altavilla                     \inst{\ref{inst:0109},\ref{inst:0110},\ref{inst:0034}}
\and M.A.      ~\'{A}lvarez                   \inst{\ref{inst:0107}}
\and R.        ~Alvarez                       \inst{\ref{inst:0014}}
\and J.        ~Alves                         \inst{\ref{inst:0105}}
\and R.I.      ~Anderson                      \inst{\ref{inst:0143},\ref{inst:0010}}
\and A.H.      ~Andrei                        \inst{\ref{inst:0145},\ref{inst:0146},\ref{inst:0058}}
\and E.        ~Anglada Varela                \inst{\ref{inst:0078}}
\and E.        ~Antiche                       \inst{\ref{inst:0012}}
\and T.        ~Antoja                        \inst{\ref{inst:0003},\ref{inst:0012}}
\and B.        ~Arcay                         \inst{\ref{inst:0107}}
\and T.L.      ~Astraatmadja                  \inst{\ref{inst:0007},\ref{inst:0154}}
\and N.        ~Bach                          \inst{\ref{inst:0097}}
\and S.G.      ~Baker                         \inst{\ref{inst:0029}}
\and L.        ~Balaguer-N\'{u}\~{n}ez        \inst{\ref{inst:0012}}
\and P.        ~Balm                          \inst{\ref{inst:0023}}
\and C.        ~Barache                       \inst{\ref{inst:0058}}
\and C.        ~Barata                        \inst{\ref{inst:0102}}
\and D.        ~Barbato                       \inst{\ref{inst:0161},\ref{inst:0030}}
\and F.        ~Barblan                       \inst{\ref{inst:0010}}
\and P.S.      ~Barklem                       \inst{\ref{inst:0101}}
\and D.        ~Barrado                       \inst{\ref{inst:0165}}
\and M.        ~Barros                        \inst{\ref{inst:0102}}
\and M.A.      ~Barstow                       \inst{\ref{inst:0167}}
\and S.        ~Bartholom\'{e} Mu\~{n}oz      \inst{\ref{inst:0012}}
\and J.-L.     ~Bassilana                     \inst{\ref{inst:0134}}
\and U.        ~Becciani                      \inst{\ref{inst:0104}}
\and M.        ~Bellazzini                    \inst{\ref{inst:0034}}
\and A.        ~Berihuete                     \inst{\ref{inst:0172}}
\and S.        ~Bertone                       \inst{\ref{inst:0030},\ref{inst:0058},\ref{inst:0175}}
\and L.        ~Bianchi                       \inst{\ref{inst:0176}}
\and O.        ~Bienaym\'{e}                  \inst{\ref{inst:0090}}
\and S.        ~Blanco-Cuaresma               \inst{\ref{inst:0010},\ref{inst:0024},\ref{inst:0180}}
\and T.        ~Boch                          \inst{\ref{inst:0090}}
\and C.        ~Boeche                        \inst{\ref{inst:0002}}
\and A.        ~Bombrun                       \inst{\ref{inst:0183}}
\and R.        ~Borrachero                    \inst{\ref{inst:0012}}
\and D.        ~Bossini                       \inst{\ref{inst:0002}}
\and S.        ~Bouquillon                    \inst{\ref{inst:0058}}
\and G.        ~Bourda                        \inst{\ref{inst:0024}}
\and A.        ~Bragaglia                     \inst{\ref{inst:0034}}
\and L.        ~Bramante                      \inst{\ref{inst:0043}}
\and M.A.      ~Breddels                      \inst{\ref{inst:0190}}
\and A.        ~Bressan                       \inst{\ref{inst:0191}}
\and N.        ~Brouillet                     \inst{\ref{inst:0024}}
\and T.        ~Br\"{ u}semeister             \inst{\ref{inst:0008}}
\and E.        ~Brugaletta                    \inst{\ref{inst:0104}}
\and B.        ~Bucciarelli                   \inst{\ref{inst:0030}}
\and A.        ~Burlacu                       \inst{\ref{inst:0018}}
\and D.        ~Busonero                      \inst{\ref{inst:0030}}
\and A.G.      ~Butkevich                     \inst{\ref{inst:0013}}
\and R.        ~Buzzi                         \inst{\ref{inst:0030}}
\and E.        ~Caffau                        \inst{\ref{inst:0005}}
\and R.        ~Cancelliere                   \inst{\ref{inst:0201}}
\and G.        ~Cannizzaro                    \inst{\ref{inst:0202},\ref{inst:0136}}
\and T.        ~Cantat-Gaudin                 \inst{\ref{inst:0002},\ref{inst:0012}}
\and R.        ~Carballo                      \inst{\ref{inst:0206}}
\and T.        ~Carlucci                      \inst{\ref{inst:0058}}
\and J.M.      ~Carrasco                      \inst{\ref{inst:0012}}
\and L.        ~Casamiquela                   \inst{\ref{inst:0012}}
\and M.        ~Castellani                    \inst{\ref{inst:0109}}
\and A.        ~Castro-Ginard                 \inst{\ref{inst:0012}}
\and P.        ~Charlot                       \inst{\ref{inst:0024}}
\and L.        ~Chemin                        \inst{\ref{inst:0213}}
\and A.        ~Chiavassa                     \inst{\ref{inst:0017}}
\and G.        ~Cocozza                       \inst{\ref{inst:0034}}
\and G.        ~Costigan                      \inst{\ref{inst:0001}}
\and S.        ~Cowell                        \inst{\ref{inst:0009}}
\and F.        ~Crifo                         \inst{\ref{inst:0005}}
\and M.        ~Crosta                        \inst{\ref{inst:0030}}
\and C.        ~Crowley                       \inst{\ref{inst:0183}}
\and J.        ~Cuypers$^\dagger$             \inst{\ref{inst:0064}}
\and C.        ~Dafonte                       \inst{\ref{inst:0107}}
\and Y.        ~Damerdji                      \inst{\ref{inst:0075},\ref{inst:0224}}
\and A.        ~Dapergolas                    \inst{\ref{inst:0061}}
\and P.        ~David                         \inst{\ref{inst:0063}}
\and M.        ~David                         \inst{\ref{inst:0227}}
\and P.        ~de Laverny                    \inst{\ref{inst:0017}}
\and F.        ~De Luise                      \inst{\ref{inst:0229}}
\and R.        ~De March                      \inst{\ref{inst:0043}}
\and D.        ~de Martino                    \inst{\ref{inst:0231}}
\and R.        ~de Souza                      \inst{\ref{inst:0232}}
\and A.        ~de Torres                     \inst{\ref{inst:0183}}
\and J.        ~Debosscher                    \inst{\ref{inst:0074}}
\and E.        ~del Pozo                      \inst{\ref{inst:0097}}
\and M.        ~Delbo                         \inst{\ref{inst:0017}}
\and A.        ~Delgado                       \inst{\ref{inst:0009}}
\and H.E.      ~Delgado                       \inst{\ref{inst:0125}}
\and P.        ~Di Matteo                     \inst{\ref{inst:0005}}
\and S.        ~Diakite                       \inst{\ref{inst:0120}}
\and C.        ~Diener                        \inst{\ref{inst:0009}}
\and E.        ~Distefano                     \inst{\ref{inst:0104}}
\and C.        ~Dolding                       \inst{\ref{inst:0029}}
\and P.        ~Drazinos                      \inst{\ref{inst:0244}}
\and J.        ~Dur\'{a}n                     \inst{\ref{inst:0116}}
\and B.        ~Edvardsson                    \inst{\ref{inst:0101}}
\and H.        ~Enke                          \inst{\ref{inst:0247}}
\and K.        ~Eriksson                      \inst{\ref{inst:0101}}
\and P.        ~Esquej                        \inst{\ref{inst:0249}}
\and G.        ~Eynard Bontemps               \inst{\ref{inst:0018}}
\and C.        ~Fabre                         \inst{\ref{inst:0251}}
\and M.        ~Fabrizio                      \inst{\ref{inst:0109},\ref{inst:0110}}
\and S.        ~Faigler                       \inst{\ref{inst:0254}}
\and A.J.      ~Falc\~{a}o                    \inst{\ref{inst:0255}}
\and M.        ~Farr\`{a}s Casas              \inst{\ref{inst:0012}}
\and L.        ~Federici                      \inst{\ref{inst:0034}}
\and G.        ~Fedorets                      \inst{\ref{inst:0114}}
\and P.        ~Fernique                      \inst{\ref{inst:0090}}
\and F.        ~Figueras                      \inst{\ref{inst:0012}}
\and F.        ~Filippi                       \inst{\ref{inst:0043}}
\and K.        ~Findeisen                     \inst{\ref{inst:0005}}
\and A.        ~Fonti                         \inst{\ref{inst:0043}}
\and E.        ~Fraile                        \inst{\ref{inst:0249}}
\and M.        ~Fraser                        \inst{\ref{inst:0009},\ref{inst:0266}}
\and B.        ~Fr\'{e}zouls                  \inst{\ref{inst:0018}}
\and M.        ~Gai                           \inst{\ref{inst:0030}}
\and S.        ~Galleti                       \inst{\ref{inst:0034}}
\and D.        ~Garabato                      \inst{\ref{inst:0107}}
\and F.        ~Garc\'{i}a-Sedano             \inst{\ref{inst:0125}}
\and A.        ~Garofalo                      \inst{\ref{inst:0272},\ref{inst:0034}}
\and N.        ~Garralda                      \inst{\ref{inst:0012}}
\and A.        ~Gavel                         \inst{\ref{inst:0101}}
\and P.        ~Gavras                        \inst{\ref{inst:0005},\ref{inst:0061},\ref{inst:0244}}
\and J.        ~Gerssen                       \inst{\ref{inst:0247}}
\and R.        ~Geyer                         \inst{\ref{inst:0013}}
\and P.        ~Giacobbe                      \inst{\ref{inst:0030}}
\and G.        ~Gilmore                       \inst{\ref{inst:0009}}
\and S.        ~Girona                        \inst{\ref{inst:0283}}
\and G.        ~Giuffrida                     \inst{\ref{inst:0110},\ref{inst:0109}}
\and F.        ~Glass                         \inst{\ref{inst:0010}}
\and M.        ~Gomes                         \inst{\ref{inst:0102}}
\and M.        ~Granvik                       \inst{\ref{inst:0114},\ref{inst:0289}}
\and A.        ~Gueguen                       \inst{\ref{inst:0005},\ref{inst:0291}}
\and A.        ~Guerrier                      \inst{\ref{inst:0134}}
\and J.        ~Guiraud                       \inst{\ref{inst:0018}}
\and R.        ~Guti\'{e}rrez-S\'{a}nchez     \inst{\ref{inst:0023}}
\and R.        ~Haigron                       \inst{\ref{inst:0005}}
\and D.        ~Hatzidimitriou                \inst{\ref{inst:0244},\ref{inst:0061}}
\and M.        ~Hauser                        \inst{\ref{inst:0008},\ref{inst:0007}}
\and M.        ~Haywood                       \inst{\ref{inst:0005}}
\and U.        ~Heiter                        \inst{\ref{inst:0101}}
\and A.        ~Helmi                         \inst{\ref{inst:0190}}
\and J.        ~Heu                           \inst{\ref{inst:0005}}
\and T.        ~Hilger                        \inst{\ref{inst:0013}}
\and D.        ~Hobbs                         \inst{\ref{inst:0015}}
\and W.        ~Hofmann                       \inst{\ref{inst:0008}}
\and G.        ~Holland                       \inst{\ref{inst:0009}}
\and H.E.      ~Huckle                        \inst{\ref{inst:0029}}
\and A.        ~Hypki                         \inst{\ref{inst:0001},\ref{inst:0310}}
\and V.        ~Icardi                        \inst{\ref{inst:0043}}
\and K.        ~Jan{\ss}en                    \inst{\ref{inst:0247}}
\and G.        ~Jevardat de Fombelle          \inst{\ref{inst:0045}}
\and P.G.      ~Jonker                        \inst{\ref{inst:0202},\ref{inst:0136}}
\and \'{A}.L.  ~Juh\'{a}sz                    \inst{\ref{inst:0316},\ref{inst:0317}}
\and F.        ~Julbe                         \inst{\ref{inst:0012}}
\and A.        ~Karampelas                    \inst{\ref{inst:0244},\ref{inst:0320}}
\and A.        ~Kewley                        \inst{\ref{inst:0009}}
\and J.        ~Klar                          \inst{\ref{inst:0247}}
\and A.        ~Kochoska                      \inst{\ref{inst:0323},\ref{inst:0324}}
\and R.        ~Kohley                        \inst{\ref{inst:0014}}
\and K.        ~Kolenberg                     \inst{\ref{inst:0326},\ref{inst:0074},\ref{inst:0180}}
\and M.        ~Kontizas                      \inst{\ref{inst:0244}}
\and E.        ~Kontizas                      \inst{\ref{inst:0061}}
\and S.E.      ~Koposov                       \inst{\ref{inst:0009},\ref{inst:0332}}
\and G.        ~Kordopatis                    \inst{\ref{inst:0017}}
\and Z.        ~Kostrzewa-Rutkowska           \inst{\ref{inst:0202},\ref{inst:0136}}
\and P.        ~Koubsky                       \inst{\ref{inst:0336}}
\and S.        ~Lambert                       \inst{\ref{inst:0058}}
\and A.F.      ~Lanza                         \inst{\ref{inst:0104}}
\and Y.        ~Lasne                         \inst{\ref{inst:0134}}
\and J.-B.     ~Lavigne                       \inst{\ref{inst:0134}}
\and Y.        ~Le Fustec                     \inst{\ref{inst:0341}}
\and C.        ~Le Poncin-Lafitte             \inst{\ref{inst:0058}}
\and Y.        ~Lebreton                      \inst{\ref{inst:0005},\ref{inst:0344}}
\and S.        ~Leccia                        \inst{\ref{inst:0231}}
\and N.        ~Leclerc                       \inst{\ref{inst:0005}}
\and I.        ~Lecoeur-Taibi                 \inst{\ref{inst:0045}}
\and H.        ~Lenhardt                      \inst{\ref{inst:0008}}
\and F.        ~Leroux                        \inst{\ref{inst:0134}}
\and S.        ~Liao                          \inst{\ref{inst:0030},\ref{inst:0351},\ref{inst:0352}}
\and E.        ~Licata                        \inst{\ref{inst:0176}}
\and H.E.P.    ~Lindstr{\o}m                  \inst{\ref{inst:0354},\ref{inst:0355}}
\and T.A.      ~Lister                        \inst{\ref{inst:0356}}
\and E.        ~Livanou                       \inst{\ref{inst:0244}}
\and A.        ~Lobel                         \inst{\ref{inst:0064}}
\and M.        ~L\'{o}pez                     \inst{\ref{inst:0165}}
\and S.        ~Managau                       \inst{\ref{inst:0134}}
\and R.G.      ~Mann                          \inst{\ref{inst:0073}}
\and G.        ~Mantelet                      \inst{\ref{inst:0008}}
\and O.        ~Marchal                       \inst{\ref{inst:0005}}
\and J.M.      ~Marchant                      \inst{\ref{inst:0364}}
\and M.        ~Marconi                       \inst{\ref{inst:0231}}
\and S.        ~Marinoni                      \inst{\ref{inst:0109},\ref{inst:0110}}
\and G.        ~Marschalk\'{o}                \inst{\ref{inst:0316},\ref{inst:0369}}
\and D.J.      ~Marshall                      \inst{\ref{inst:0370}}
\and M.        ~Martino                       \inst{\ref{inst:0043}}
\and G.        ~Marton                        \inst{\ref{inst:0316}}
\and N.        ~Mary                          \inst{\ref{inst:0134}}
\and D.        ~Massari                       \inst{\ref{inst:0190}}
\and G.        ~Matijevi\v{c}                 \inst{\ref{inst:0247}}
\and T.        ~Mazeh                         \inst{\ref{inst:0254}}
\and P.J.      ~McMillan                      \inst{\ref{inst:0015}}
\and S.        ~Messina                       \inst{\ref{inst:0104}}
\and D.        ~Michalik                      \inst{\ref{inst:0015}}
\and N.R.      ~Millar                        \inst{\ref{inst:0009}}
\and D.        ~Molina                        \inst{\ref{inst:0012}}
\and R.        ~Molinaro                      \inst{\ref{inst:0231}}
\and L.        ~Moln\'{a}r                    \inst{\ref{inst:0316}}
\and P.        ~Montegriffo                   \inst{\ref{inst:0034}}
\and R.        ~Mor                           \inst{\ref{inst:0012}}
\and R.        ~Morbidelli                    \inst{\ref{inst:0030}}
\and T.        ~Morel                         \inst{\ref{inst:0075}}
\and D.        ~Morris                        \inst{\ref{inst:0073}}
\and A.F.      ~Mulone                        \inst{\ref{inst:0043}}
\and T.        ~Muraveva                      \inst{\ref{inst:0034}}
\and I.        ~Musella                       \inst{\ref{inst:0231}}
\and G.        ~Nelemans                      \inst{\ref{inst:0136},\ref{inst:0074}}
\and L.        ~Nicastro                      \inst{\ref{inst:0034}}
\and L.        ~Noval                         \inst{\ref{inst:0134}}
\and W.        ~O'Mullane                     \inst{\ref{inst:0014},\ref{inst:0089}}
\and C.        ~Ord\'{e}novic                 \inst{\ref{inst:0017}}
\and D.        ~Ord\'{o}\~{n}ez-Blanco        \inst{\ref{inst:0045}}
\and P.        ~Osborne                       \inst{\ref{inst:0009}}
\and C.        ~Pagani                        \inst{\ref{inst:0167}}
\and I.        ~Pagano                        \inst{\ref{inst:0104}}
\and F.        ~Pailler                       \inst{\ref{inst:0018}}
\and H.        ~Palacin                       \inst{\ref{inst:0134}}
\and L.        ~Palaversa                     \inst{\ref{inst:0009},\ref{inst:0010}}
\and A.        ~Panahi                        \inst{\ref{inst:0254}}
\and M.        ~Pawlak                        \inst{\ref{inst:0408},\ref{inst:0409}}
\and A.M.      ~Piersimoni                    \inst{\ref{inst:0229}}
\and F.-X.     ~Pineau                        \inst{\ref{inst:0090}}
\and E.        ~Plachy                        \inst{\ref{inst:0316}}
\and G.        ~Plum                          \inst{\ref{inst:0005}}
\and E.        ~Poggio                        \inst{\ref{inst:0161},\ref{inst:0030}}
\and E.        ~Poujoulet                     \inst{\ref{inst:0416}}
\and A.        ~Pr\v{s}a                      \inst{\ref{inst:0324}}
\and L.        ~Pulone                        \inst{\ref{inst:0109}}
\and E.        ~Racero                        \inst{\ref{inst:0037}}
\and S.        ~Ragaini                       \inst{\ref{inst:0034}}
\and N.        ~Rambaux                       \inst{\ref{inst:0063}}
\and M.        ~Ramos-Lerate                  \inst{\ref{inst:0422}}
\and S.        ~Regibo                        \inst{\ref{inst:0074}}
\and C.        ~Reyl\'{e}                     \inst{\ref{inst:0120}}
\and F.        ~Riclet                        \inst{\ref{inst:0018}}
\and V.        ~Ripepi                        \inst{\ref{inst:0231}}
\and A.        ~Riva                          \inst{\ref{inst:0030}}
\and A.        ~Rivard                        \inst{\ref{inst:0134}}
\and G.        ~Rixon                         \inst{\ref{inst:0009}}
\and T.        ~Roegiers                      \inst{\ref{inst:0430}}
\and M.        ~Roelens                       \inst{\ref{inst:0010}}
\and M.        ~Romero-G\'{o}mez              \inst{\ref{inst:0012}}
\and N.        ~Rowell                        \inst{\ref{inst:0073}}
\and F.        ~Royer                         \inst{\ref{inst:0005}}
\and L.        ~Ruiz-Dern                     \inst{\ref{inst:0005}}
\and G.        ~Sadowski                      \inst{\ref{inst:0019}}
\and T.        ~Sagrist\`{a} Sell\'{e}s       \inst{\ref{inst:0008}}
\and J.        ~Sahlmann                      \inst{\ref{inst:0014},\ref{inst:0439}}
\and J.        ~Salgado                       \inst{\ref{inst:0440}}
\and E.        ~Salguero                      \inst{\ref{inst:0078}}
\and N.        ~Sanna                         \inst{\ref{inst:0021}}
\and T.        ~Santana-Ros                   \inst{\ref{inst:0310}}
\and M.        ~Sarasso                       \inst{\ref{inst:0030}}
\and H.        ~Savietto                      \inst{\ref{inst:0445}}
\and M.        ~Schultheis                    \inst{\ref{inst:0017}}
\and E.        ~Sciacca                       \inst{\ref{inst:0104}}
\and M.        ~Segol                         \inst{\ref{inst:0448}}
\and J.C.      ~Segovia                       \inst{\ref{inst:0037}}
\and D.        ~S\'{e}gransan                 \inst{\ref{inst:0010}}
\and I-C.      ~Shih                          \inst{\ref{inst:0005}}
\and L.        ~Siltala                       \inst{\ref{inst:0114},\ref{inst:0453}}
\and A.F.      ~Silva                         \inst{\ref{inst:0102}}
\and R.L.      ~Smart                         \inst{\ref{inst:0030}}
\and K.W.      ~Smith                         \inst{\ref{inst:0007}}
\and E.        ~Solano                        \inst{\ref{inst:0165},\ref{inst:0458}}
\and F.        ~Solitro                       \inst{\ref{inst:0043}}
\and R.        ~Sordo                         \inst{\ref{inst:0002}}
\and S.        ~Soria Nieto                   \inst{\ref{inst:0012}}
\and J.        ~Souchay                       \inst{\ref{inst:0058}}
\and A.        ~Spagna                        \inst{\ref{inst:0030}}
\and F.        ~Spoto                         \inst{\ref{inst:0017},\ref{inst:0063}}
\and U.        ~Stampa                        \inst{\ref{inst:0008}}
\and I.A.      ~Steele                        \inst{\ref{inst:0364}}
\and H.        ~Steidelm\"{ u}ller            \inst{\ref{inst:0013}}
\and C.A.      ~Stephenson                    \inst{\ref{inst:0023}}
\and H.        ~Stoev                         \inst{\ref{inst:0470}}
\and F.F.      ~Suess                         \inst{\ref{inst:0009}}
\and J.        ~Surdej                        \inst{\ref{inst:0075}}
\and L.        ~Szabados                      \inst{\ref{inst:0316}}
\and E.        ~Szegedi-Elek                  \inst{\ref{inst:0316}}
\and D.        ~Tapiador                      \inst{\ref{inst:0475},\ref{inst:0476}}
\and F.        ~Taris                         \inst{\ref{inst:0058}}
\and G.        ~Tauran                        \inst{\ref{inst:0134}}
\and M.B.      ~Taylor                        \inst{\ref{inst:0479}}
\and R.        ~Teixeira                      \inst{\ref{inst:0232}}
\and D.        ~Terrett                       \inst{\ref{inst:0122}}
\and P.        ~Teyssandier                   \inst{\ref{inst:0058}}
\and W.        ~Thuillot                      \inst{\ref{inst:0063}}
\and A.        ~Titarenko                     \inst{\ref{inst:0017}}
\and F.        ~Torra Clotet                  \inst{\ref{inst:0485}}
\and C.        ~Turon                         \inst{\ref{inst:0005}}
\and A.        ~Ulla                          \inst{\ref{inst:0487}}
\and E.        ~Utrilla                       \inst{\ref{inst:0097}}
\and S.        ~Uzzi                          \inst{\ref{inst:0043}}
\and M.        ~Vaillant                      \inst{\ref{inst:0134}}
\and G.        ~Valentini                     \inst{\ref{inst:0229}}
\and V.        ~Valette                       \inst{\ref{inst:0018}}
\and A.        ~van Elteren                   \inst{\ref{inst:0001}}
\and E.        ~Van Hemelryck                 \inst{\ref{inst:0064}}
\and M.        ~van Leeuwen                   \inst{\ref{inst:0009}}
\and M.        ~Vaschetto                     \inst{\ref{inst:0043}}
\and A.        ~Vecchiato                     \inst{\ref{inst:0030}}
\and J.        ~Veljanoski                    \inst{\ref{inst:0190}}
\and Y.        ~Viala                         \inst{\ref{inst:0005}}
\and D.        ~Vicente                       \inst{\ref{inst:0283}}
\and S.        ~Vogt                          \inst{\ref{inst:0430}}
\and C.        ~von Essen                     \inst{\ref{inst:0502}}
\and H.        ~Voss                          \inst{\ref{inst:0012}}
\and V.        ~Votruba                       \inst{\ref{inst:0336}}
\and S.        ~Voutsinas                     \inst{\ref{inst:0073}}
\and G.        ~Walmsley                      \inst{\ref{inst:0018}}
\and M.        ~Weiler                        \inst{\ref{inst:0012}}
\and O.        ~Wertz                         \inst{\ref{inst:0508}}
\and T.        ~Wevers                        \inst{\ref{inst:0009},\ref{inst:0136}}
\and \L{}.     ~Wyrzykowski                   \inst{\ref{inst:0009},\ref{inst:0408}}
\and A.        ~Yoldas                        \inst{\ref{inst:0009}}
\and M.        ~\v{Z}erjal                    \inst{\ref{inst:0323},\ref{inst:0515}}
\and H.        ~Ziaeepour                     \inst{\ref{inst:0120}}
\and J.        ~Zorec                         \inst{\ref{inst:0517}}
\and S.        ~Zschocke                      \inst{\ref{inst:0013}}
\and S.        ~Zucker                        \inst{\ref{inst:0519}}
\and C.        ~Zurbach                       \inst{\ref{inst:0098}}
\and T.        ~Zwitter                       \inst{\ref{inst:0323}}
}
\institute{
     Leiden Observatory, Leiden University, Niels Bohrweg 2, 2333 CA Leiden, The Netherlands\relax                                                                                                           \label{inst:0001}
\and INAF - Osservatorio astronomico di Padova, Vicolo Osservatorio 5, 35122 Padova, Italy\relax                                                                                                             \label{inst:0002}
\and Science Support Office, Directorate of Science, European Space Research and Technology Centre (ESA/ESTEC), Keplerlaan 1, 2201AZ, Noordwijk, The Netherlands\relax                                       \label{inst:0003}
\and GEPI, Observatoire de Paris, Universit\'{e} PSL, CNRS, 5 Place Jules Janssen, 92190 Meudon, France\relax                                                                                                \label{inst:0005}
\and Univ. Grenoble Alpes, CNRS, IPAG, 38000 Grenoble, France\relax                                                                                                                                          \label{inst:0006}
\and Max Planck Institute for Astronomy, K\"{ o}nigstuhl 17, 69117 Heidelberg, Germany\relax                                                                                                                 \label{inst:0007}
\and Astronomisches Rechen-Institut, Zentrum f\"{ u}r Astronomie der Universit\"{ a}t Heidelberg, M\"{ o}nchhofstr. 12-14, 69120 Heidelberg, Germany\relax                                                   \label{inst:0008}
\and Institute of Astronomy, University of Cambridge, Madingley Road, Cambridge CB3 0HA, United Kingdom\relax                                                                                                \label{inst:0009}
\and Department of Astronomy, University of Geneva, Chemin des Maillettes 51, 1290 Versoix, Switzerland\relax                                                                                                \label{inst:0010}
\and Mission Operations Division, Operations Department, Directorate of Science, European Space Research and Technology Centre (ESA/ESTEC), Keplerlaan 1, 2201 AZ, Noordwijk, The Netherlands\relax          \label{inst:0011}
\and Institut de Ci\`{e}ncies del Cosmos, Universitat  de  Barcelona  (IEEC-UB), Mart\'{i} i  Franqu\`{e}s  1, 08028 Barcelona, Spain\relax                                                                  \label{inst:0012}
\and Lohrmann Observatory, Technische Universit\"{ a}t Dresden, Mommsenstra{\ss}e 13, 01062 Dresden, Germany\relax                                                                                           \label{inst:0013}
\and European Space Astronomy Centre (ESA/ESAC), Camino bajo del Castillo, s/n, Urbanizacion Villafranca del Castillo, Villanueva de la Ca\~{n}ada, 28692 Madrid, Spain\relax                                \label{inst:0014}
\and Lund Observatory, Department of Astronomy and Theoretical Physics, Lund University, Box 43, 22100 Lund, Sweden\relax                                                                                    \label{inst:0015}
\and Universit\'{e} C\^{o}te d'Azur, Observatoire de la C\^{o}te d'Azur, CNRS, Laboratoire Lagrange, Bd de l'Observatoire, CS 34229, 06304 Nice Cedex 4, France\relax                                        \label{inst:0017}
\and CNES Centre Spatial de Toulouse, 18 avenue Edouard Belin, 31401 Toulouse Cedex 9, France\relax                                                                                                          \label{inst:0018}
\and Institut d'Astronomie et d'Astrophysique, Universit\'{e} Libre de Bruxelles CP 226, Boulevard du Triomphe, 1050 Brussels, Belgium\relax                                                                 \label{inst:0019}
\and F.R.S.-FNRS, Rue d'Egmont 5, 1000 Brussels, Belgium\relax                                                                                                                                               \label{inst:0020}
\and INAF - Osservatorio Astrofisico di Arcetri, Largo Enrico Fermi 5, 50125 Firenze, Italy\relax                                                                                                            \label{inst:0021}
\and Telespazio Vega UK Ltd for ESA/ESAC, Camino bajo del Castillo, s/n, Urbanizacion Villafranca del Castillo, Villanueva de la Ca\~{n}ada, 28692 Madrid, Spain\relax                                       \label{inst:0023}
\and Laboratoire d'astrophysique de Bordeaux, Univ. Bordeaux, CNRS, B18N, all{\'e}e Geoffroy Saint-Hilaire, 33615 Pessac, France\relax                                                                       \label{inst:0024}
\and Mullard Space Science Laboratory, University College London, Holmbury St Mary, Dorking, Surrey RH5 6NT, United Kingdom\relax                                                                            \label{inst:0029}
\and INAF - Osservatorio Astrofisico di Torino, via Osservatorio 20, 10025 Pino Torinese (TO), Italy\relax                                                                                                   \label{inst:0030}
\and INAF - Osservatorio di Astrofisica e Scienza dello Spazio di Bologna, via Piero Gobetti 93/3, 40129 Bologna, Italy\relax                                                                                \label{inst:0034}
\and Serco Gesti\'{o}n de Negocios for ESA/ESAC, Camino bajo del Castillo, s/n, Urbanizacion Villafranca del Castillo, Villanueva de la Ca\~{n}ada, 28692 Madrid, Spain\relax                                \label{inst:0037}
\and ALTEC S.p.a, Corso Marche, 79,10146 Torino, Italy\relax                                                                                                                                                 \label{inst:0043}
\and Department of Astronomy, University of Geneva, Chemin d'Ecogia 16, 1290 Versoix, Switzerland\relax                                                                                                      \label{inst:0045}
\and Gaia DPAC Project Office, ESAC, Camino bajo del Castillo, s/n, Urbanizacion Villafranca del Castillo, Villanueva de la Ca\~{n}ada, 28692 Madrid, Spain\relax                                            \label{inst:0052}
\and SYRTE, Observatoire de Paris, Universit\'{e} PSL, CNRS,  Sorbonne Universit\'{e}, LNE, 61 avenue de l’Observatoire 75014 Paris, France\relax                                                          \label{inst:0058}
\and National Observatory of Athens, I. Metaxa and Vas. Pavlou, Palaia Penteli, 15236 Athens, Greece\relax                                                                                                   \label{inst:0061}
\and IMCCE, Observatoire de Paris, Universit\'{e} PSL, CNRS,  Sorbonne Universit\'{e}, Univ. Lille, 77 av. Denfert-Rochereau, 75014 Paris, France\relax                                                      \label{inst:0063}
\and Royal Observatory of Belgium, Ringlaan 3, 1180 Brussels, Belgium\relax                                                                                                                                  \label{inst:0064}
\and Institute for Astronomy, University of Edinburgh, Royal Observatory, Blackford Hill, Edinburgh EH9 3HJ, United Kingdom\relax                                                                            \label{inst:0073}
\and Instituut voor Sterrenkunde, KU Leuven, Celestijnenlaan 200D, 3001 Leuven, Belgium\relax                                                                                                                \label{inst:0074}
\and Institut d'Astrophysique et de G\'{e}ophysique, Universit\'{e} de Li\`{e}ge, 19c, All\'{e}e du 6 Ao\^{u}t, B-4000 Li\`{e}ge, Belgium\relax                                                              \label{inst:0075}
\and ATG Europe for ESA/ESAC, Camino bajo del Castillo, s/n, Urbanizacion Villafranca del Castillo, Villanueva de la Ca\~{n}ada, 28692 Madrid, Spain\relax                                                   \label{inst:0078}
\and \'{A}rea de Lenguajes y Sistemas Inform\'{a}ticos, Universidad Pablo de Olavide, Ctra. de Utrera, km 1. 41013, Sevilla, Spain\relax                                                                     \label{inst:0082}
\and ETSE Telecomunicaci\'{o}n, Universidade de Vigo, Campus Lagoas-Marcosende, 36310 Vigo, Galicia, Spain\relax                                                                                             \label{inst:0084}
\and Large Synoptic Survey Telescope, 950 N. Cherry Avenue, Tucson, AZ 85719, USA\relax                                                                                                                      \label{inst:0089}
\and Observatoire Astronomique de Strasbourg, Universit\'{e} de Strasbourg, CNRS, UMR 7550, 11 rue de l'Universit\'{e}, 67000 Strasbourg, France\relax                                                       \label{inst:0090}
\and Kavli Institute for Cosmology, University of Cambridge, Madingley Road, Cambride CB3 0HA, United Kingdom\relax                                                                                          \label{inst:0093}
\and Aurora Technology for ESA/ESAC, Camino bajo del Castillo, s/n, Urbanizacion Villafranca del Castillo, Villanueva de la Ca\~{n}ada, 28692 Madrid, Spain\relax                                            \label{inst:0097}
\and Laboratoire Univers et Particules de Montpellier, Universit\'{e} Montpellier, Place Eug\`{e}ne Bataillon, CC72, 34095 Montpellier Cedex 05, France\relax                                                \label{inst:0098}
\and Department of Physics and Astronomy, Division of Astronomy and Space Physics, Uppsala University, Box 516, 75120 Uppsala, Sweden\relax                                                                  \label{inst:0101}
\and CENTRA, Universidade de Lisboa, FCUL, Campo Grande, Edif. C8, 1749-016 Lisboa, Portugal\relax                                                                                                           \label{inst:0102}
\and Universit\`{a} di Catania, Dipartimento di Fisica e Astronomia, Sezione Astrofisica, Via S. Sofia 78, 95123 Catania, Italy\relax                                                                        \label{inst:0103}
\and INAF - Osservatorio Astrofisico di Catania, via S. Sofia 78, 95123 Catania, Italy\relax                                                                                                                 \label{inst:0104}
\and University of Vienna, Department of Astrophysics, T\"{ u}rkenschanzstra{\ss}e 17, A1180 Vienna, Austria\relax                                                                                           \label{inst:0105}
\and CITIC – Department of Computer Science, University of A Coru\~{n}a, Campus de Elvi\~{n}a S/N, 15071, A Coru\~{n}a, Spain\relax                                                                        \label{inst:0107}
\and CITIC – Astronomy and Astrophysics, University of A Coru\~{n}a, Campus de Elvi\~{n}a S/N, 15071, A Coru\~{n}a, Spain\relax                                                                            \label{inst:0108}
\and INAF - Osservatorio Astronomico di Roma, Via di Frascati 33, 00078 Monte Porzio Catone (Roma), Italy\relax                                                                                              \label{inst:0109}
\and Space Science Data Center - ASI, Via del Politecnico SNC, 00133 Roma, Italy\relax                                                                                                                       \label{inst:0110}
\and University of Helsinki, Department of Physics, P.O. Box 64, 00014 Helsinki, Finland\relax                                                                                                               \label{inst:0114}
\and Finnish Geospatial Research Institute FGI, Geodeetinrinne 2, 02430 Masala, Finland\relax                                                                                                                \label{inst:0115}
\and Isdefe for ESA/ESAC, Camino bajo del Castillo, s/n, Urbanizacion Villafranca del Castillo, Villanueva de la Ca\~{n}ada, 28692 Madrid, Spain\relax                                                       \label{inst:0116}
\and Institut UTINAM UMR6213, CNRS, OSU THETA Franche-Comt\'{e} Bourgogne, Universit\'{e} Bourgogne Franche-Comt\'{e}, 25000 Besan\c{c}on, France\relax                                                      \label{inst:0120}
\and STFC, Rutherford Appleton Laboratory, Harwell, Didcot, OX11 0QX, United Kingdom\relax                                                                                                                   \label{inst:0122}
\and Dpto. de Inteligencia Artificial, UNED, c/ Juan del Rosal 16, 28040 Madrid, Spain\relax                                                                                                                 \label{inst:0125}
\and Elecnor Deimos Space for ESA/ESAC, Camino bajo del Castillo, s/n, Urbanizacion Villafranca del Castillo, Villanueva de la Ca\~{n}ada, 28692 Madrid, Spain\relax                                         \label{inst:0133}
\and Thales Services for CNES Centre Spatial de Toulouse, 18 avenue Edouard Belin, 31401 Toulouse Cedex 9, France\relax                                                                                      \label{inst:0134}
\and Department of Astrophysics/IMAPP, Radboud University, P.O.Box 9010, 6500 GL Nijmegen, The Netherlands\relax                                                                                             \label{inst:0136}
\and European Southern Observatory, Karl-Schwarzschild-Str. 2, 85748 Garching, Germany\relax                                                                                                                 \label{inst:0143}
\and ON/MCTI-BR, Rua Gal. Jos\'{e} Cristino 77, Rio de Janeiro, CEP 20921-400, RJ,  Brazil\relax                                                                                                             \label{inst:0145}
\and OV/UFRJ-BR, Ladeira Pedro Ant\^{o}nio 43, Rio de Janeiro, CEP 20080-090, RJ, Brazil\relax                                                                                                               \label{inst:0146}
\and Department of Terrestrial Magnetism, Carnegie Institution for Science, 5241 Broad Branch Road, NW, Washington, DC 20015-1305, USA\relax                                                                 \label{inst:0154}
\and Universit\`{a} di Torino, Dipartimento di Fisica, via Pietro Giuria 1, 10125 Torino, Italy\relax                                                                                                        \label{inst:0161}
\and Departamento de Astrof\'{i}sica, Centro de Astrobiolog\'{i}a (CSIC-INTA), ESA-ESAC. Camino Bajo del Castillo s/n. 28692 Villanueva de la Ca\~{n}ada, Madrid, Spain\relax                                \label{inst:0165}
\and Leicester Institute of Space and Earth Observation and Department of Physics and Astronomy, University of Leicester, University Road, Leicester LE1 7RH, United Kingdom\relax                           \label{inst:0167}
\and Departamento de Estad\'{i}stica, Universidad de C\'{a}diz, Calle Rep\'{u}blica \'{A}rabe Saharawi s/n. 11510, Puerto Real, C\'{a}diz, Spain\relax                                                       \label{inst:0172}
\and Astronomical Institute Bern University, Sidlerstrasse 5, 3012 Bern, Switzerland (present address)\relax                                                                                                 \label{inst:0175}
\and EURIX S.r.l., Corso Vittorio Emanuele II 61, 10128, Torino, Italy\relax                                                                                                                                 \label{inst:0176}
\and Harvard-Smithsonian Center for Astrophysics, 60 Garden Street, Cambridge MA 02138, USA\relax                                                                                                            \label{inst:0180}
\and HE Space Operations BV for ESA/ESAC, Camino bajo del Castillo, s/n, Urbanizacion Villafranca del Castillo, Villanueva de la Ca\~{n}ada, 28692 Madrid, Spain\relax                                       \label{inst:0183}
\and Kapteyn Astronomical Institute, University of Groningen, Landleven 12, 9747 AD Groningen, The Netherlands\relax                                                                                         \label{inst:0190}
\and SISSA - Scuola Internazionale Superiore di Studi Avanzati, via Bonomea 265, 34136 Trieste, Italy\relax                                                                                                  \label{inst:0191}
\and University of Turin, Department of Computer Sciences, Corso Svizzera 185, 10149 Torino, Italy\relax                                                                                                     \label{inst:0201}
\and SRON, Netherlands Institute for Space Research, Sorbonnelaan 2, 3584CA, Utrecht, The Netherlands\relax                                                                                                  \label{inst:0202}
\and Dpto. de Matem\'{a}tica Aplicada y Ciencias de la Computaci\'{o}n, Univ. de Cantabria, ETS Ingenieros de Caminos, Canales y Puertos, Avda. de los Castros s/n, 39005 Santander, Spain\relax             \label{inst:0206}
\and Unidad de Astronom\'ia, Universidad de Antofagasta, Avenida Angamos 601, Antofagasta 1270300, Chile\relax                                                                                               \label{inst:0213}
\and CRAAG - Centre de Recherche en Astronomie, Astrophysique et G\'{e}ophysique, Route de l'Observatoire Bp 63 Bouzareah 16340 Algiers, Algeria\relax                                                       \label{inst:0224}
\and University of Antwerp, Onderzoeksgroep Toegepaste Wiskunde, Middelheimlaan 1, 2020 Antwerp, Belgium\relax                                                                                               \label{inst:0227}
\and INAF - Osservatorio Astronomico d'Abruzzo, Via Mentore Maggini, 64100 Teramo, Italy\relax                                                                                                               \label{inst:0229}
\and INAF - Osservatorio Astronomico di Capodimonte, Via Moiariello 16, 80131, Napoli, Italy\relax                                                                                                           \label{inst:0231}
\and Instituto de Astronomia, Geof\`{i}sica e Ci\^{e}ncias Atmosf\'{e}ricas, Universidade de S\~{a}o Paulo, Rua do Mat\~{a}o, 1226, Cidade Universitaria, 05508-900 S\~{a}o Paulo, SP, Brazil\relax          \label{inst:0232}
\and Department of Astrophysics, Astronomy and Mechanics, National and Kapodistrian University of Athens, Panepistimiopolis, Zografos, 15783 Athens, Greece\relax                                            \label{inst:0244}
\and Leibniz Institute for Astrophysics Potsdam (AIP), An der Sternwarte 16, 14482 Potsdam, Germany\relax                                                                                                    \label{inst:0247}
\and RHEA for ESA/ESAC, Camino bajo del Castillo, s/n, Urbanizacion Villafranca del Castillo, Villanueva de la Ca\~{n}ada, 28692 Madrid, Spain\relax                                                         \label{inst:0249}
\and ATOS for CNES Centre Spatial de Toulouse, 18 avenue Edouard Belin, 31401 Toulouse Cedex 9, France\relax                                                                                                 \label{inst:0251}
\and School of Physics and Astronomy, Tel Aviv University, Tel Aviv 6997801, Israel\relax                                                                                                                    \label{inst:0254}
\and UNINOVA - CTS, Campus FCT-UNL, Monte da Caparica, 2829-516 Caparica, Portugal\relax                                                                                                                     \label{inst:0255}
\and School of Physics, O'Brien Centre for Science North, University College Dublin, Belfield, Dublin 4, Ireland\relax                                                                                       \label{inst:0266}
\and Dipartimento di Fisica e Astronomia, Universit\`{a} di Bologna, Via Piero Gobetti 93/2, 40129 Bologna, Italy\relax                                                                                      \label{inst:0272}
\and Barcelona Supercomputing Center - Centro Nacional de Supercomputaci\'{o}n, c/ Jordi Girona 29, Ed. Nexus II, 08034 Barcelona, Spain\relax                                                               \label{inst:0283}
\and Department of Computer Science, Electrical and Space Engineering, Lule\aa{} University of Technology, Box 848, S-981 28 Kiruna, Sweden\relax                                                            \label{inst:0289}
\and Max Planck Institute for Extraterrestrial Physics, High Energy Group, Gie{\ss}enbachstra{\ss}e, 85741 Garching, Germany\relax                                                                           \label{inst:0291}
\and Astronomical Observatory Institute, Faculty of Physics, Adam Mickiewicz University, S{\l}oneczna 36, 60-286 Pozna{\'n}, Poland\relax                                                                    \label{inst:0310}
\and Konkoly Observatory, Research Centre for Astronomy and Earth Sciences, Hungarian Academy of Sciences, Konkoly Thege Mikl\'{o}s \'{u}t 15-17, 1121 Budapest, Hungary\relax                               \label{inst:0316}
\and E\"{ o}tv\"{ o}s Lor\'and University, Egyetem t\'{e}r 1-3, H-1053 Budapest, Hungary\relax                                                                                                               \label{inst:0317}
\and American Community Schools of Athens, 129 Aghias Paraskevis Ave. \& Kazantzaki Street, Halandri, 15234 Athens, Greece\relax                                                                             \label{inst:0320}
\and Faculty of Mathematics and Physics, University of Ljubljana, Jadranska ulica 19, 1000 Ljubljana, Slovenia\relax                                                                                         \label{inst:0323}
\and Villanova University, Department of Astrophysics and Planetary Science, 800 E Lancaster Avenue, Villanova PA 19085, USA\relax                                                                           \label{inst:0324}
\and Physics Department, University of Antwerp, Groenenborgerlaan 171, 2020 Antwerp, Belgium\relax                                                                                                           \label{inst:0326}
\and McWilliams Center for Cosmology, Department of Physics, Carnegie Mellon University, 5000 Forbes Avenue, Pittsburgh, PA 15213, USA\relax                                                                 \label{inst:0332}
\and Astronomical Institute, Academy of Sciences of the Czech Republic, Fri\v{c}ova 298, 25165 Ond\v{r}ejov, Czech Republic\relax                                                                            \label{inst:0336}
\and Telespazio for CNES Centre Spatial de Toulouse, 18 avenue Edouard Belin, 31401 Toulouse Cedex 9, France\relax                                                                                           \label{inst:0341}
\and Institut de Physique de Rennes, Universit{\'e} de Rennes 1, 35042 Rennes, France\relax                                                                                                                  \label{inst:0344}
\and Shanghai Astronomical Observatory, Chinese Academy of Sciences, 80 Nandan Rd, 200030 Shanghai, China\relax                                                                                              \label{inst:0351}
\and School of Astronomy and Space Science, University of Chinese Academy of Sciences, Beijing 100049, China\relax                                                                                           \label{inst:0352}
\and Niels Bohr Institute, University of Copenhagen, Juliane Maries Vej 30, 2100 Copenhagen {\O}, Denmark\relax                                                                                              \label{inst:0354}
\and DXC Technology, Retortvej 8, 2500 Valby, Denmark\relax                                                                                                                                                  \label{inst:0355}
\and Las Cumbres Observatory, 6740 Cortona Drive Suite 102, Goleta, CA 93117, USA\relax                                                                                                                      \label{inst:0356}
\and Astrophysics Research Institute, Liverpool John Moores University, 146 Brownlow Hill, Liverpool L3 5RF, United Kingdom\relax                                                                            \label{inst:0364}
\and Baja Observatory of University of Szeged, Szegedi \'{u}t III/70, 6500 Baja, Hungary\relax                                                                                                               \label{inst:0369}
\and Laboratoire AIM, IRFU/Service d'Astrophysique - CEA/DSM - CNRS - Universit\'{e} Paris Diderot, B\^{a}t 709, CEA-Saclay, 91191 Gif-sur-Yvette Cedex, France\relax                                        \label{inst:0370}
\and Warsaw University Observatory, Al. Ujazdowskie 4, 00-478 Warszawa, Poland\relax                                                                                                                         \label{inst:0408}
\and Institute of Theoretical Physics, Faculty of Mathematics and Physics, Charles University in Prague, Czech Republic\relax                                                                                \label{inst:0409}
\and AKKA for CNES Centre Spatial de Toulouse, 18 avenue Edouard Belin, 31401 Toulouse Cedex 9, France\relax                                                                                                 \label{inst:0416}
\and Vitrociset Belgium for ESA/ESAC, Camino bajo del Castillo, s/n, Urbanizacion Villafranca del Castillo, Villanueva de la Ca\~{n}ada, 28692 Madrid, Spain\relax                                           \label{inst:0422}
\and HE Space Operations BV for ESA/ESTEC, Keplerlaan 1, 2201AZ, Noordwijk, The Netherlands\relax                                                                                                            \label{inst:0430}
\and Space Telescope Science Institute, 3700 San Martin Drive, Baltimore, MD 21218, USA\relax                                                                                                                \label{inst:0439}
\and QUASAR Science Resources for ESA/ESAC, Camino bajo del Castillo, s/n, Urbanizacion Villafranca del Castillo, Villanueva de la Ca\~{n}ada, 28692 Madrid, Spain\relax                                     \label{inst:0440}
\and Fork Research, Rua do Cruzado Osberno, Lt. 1, 9 esq., Lisboa, Portugal\relax                                                                                                                            \label{inst:0445}
\and APAVE SUDEUROPE SAS for CNES Centre Spatial de Toulouse, 18 avenue Edouard Belin, 31401 Toulouse Cedex 9, France\relax                                                                                  \label{inst:0448}
\and Nordic Optical Telescope, Rambla Jos\'{e} Ana Fern\'{a}ndez P\'{e}rez 7, 38711 Bre\~{n}a Baja, Spain\relax                                                                                              \label{inst:0453}
\and Spanish Virtual Observatory\relax                                                                                                                                                                       \label{inst:0458}
\and Fundaci\'{o}n Galileo Galilei - INAF, Rambla Jos\'{e} Ana Fern\'{a}ndez P\'{e}rez 7, 38712 Bre\~{n}a Baja, Santa Cruz de Tenerife, Spain\relax                                                          \label{inst:0470}
\and INSA for ESA/ESAC, Camino bajo del Castillo, s/n, Urbanizacion Villafranca del Castillo, Villanueva de la Ca\~{n}ada, 28692 Madrid, Spain\relax                                                         \label{inst:0475}
\and Dpto. Arquitectura de Computadores y Autom\'{a}tica, Facultad de Inform\'{a}tica, Universidad Complutense de Madrid, C/ Prof. Jos\'{e} Garc\'{i}a Santesmases s/n, 28040 Madrid, Spain\relax            \label{inst:0476}
\and H H Wills Physics Laboratory, University of Bristol, Tyndall Avenue, Bristol BS8 1TL, United Kingdom\relax                                                                                              \label{inst:0479}
\and Institut d'Estudis Espacials de Catalunya (IEEC), Gran Capita 2-4, 08034 Barcelona, Spain\relax                                                                                                         \label{inst:0485}
\and Applied Physics Department, Universidade de Vigo, 36310 Vigo, Spain\relax                                                                                                                               \label{inst:0487}
\and Stellar Astrophysics Centre, Aarhus University, Department of Physics and Astronomy, 120 Ny Munkegade, Building 1520, DK-8000 Aarhus C, Denmark\relax                                                   \label{inst:0502}
\and Argelander-Institut f\"{ ur} Astronomie, Universit\"{ a}t Bonn,  Auf dem H\"{ u}gel 71, 53121 Bonn, Germany\relax                                                                                       \label{inst:0508}
\and Research School of Astronomy and Astrophysics, Australian National University, Canberra, ACT 2611 Australia\relax                                                                                       \label{inst:0515}
\and Sorbonne Universit\'{e}s, UPMC Univ. Paris 6 et CNRS, UMR 7095, Institut d'Astrophysique de Paris, 98 bis bd. Arago, 75014 Paris, France\relax                                                          \label{inst:0517}
\and Department of Geosciences, Tel Aviv University, Tel Aviv 6997801, Israel\relax                                                                                                                          \label{inst:0519}
}

\date{Received ; accepted }

\abstract{We present the second {\gaia} data release, \gdr{2}, consisting of astrometry,
photometry, radial velocities, and information on astrophysical parameters and variability, for
sources brighter than magnitude $21$. In addition epoch astrometry and photometry are provided for a
modest sample of minor planets in the solar system.}
{A summary of the contents of \gdr{2} is presented, accompanied by a discussion on the
differences with respect to \gdr{1} and an overview of the main limitations which are still
present in the survey. Recommendations are made on the responsible use of \gdr{2} results.}
{The raw data collected with the {\gaia} instruments during the first 22 months of the mission have
been processed by the {\gaia} Data Processing and Analysis Consortium (DPAC) and turned into this
second data release, which represents a major advance with respect to \gdr{1} in terms of
completeness, performance, and richness of the data products.}
{\gdr{2} contains celestial positions and the apparent brightness in $G$ for approximately $1.7$
billion sources. For $1.3$ billion of those sources, parallaxes and proper motions are in addition
available. The sample of sources for which variability information is provided is expanded to $0.5$
million stars. This data release contains four new elements: broad-band colour information in the form of
the apparent brightness in the {\gbp} (330--680~nm) and {\grp} (630--1050~nm) bands is available for
$1.4$ billion sources; median radial velocities for some 7 million sources are presented; for
between $77$ and $161$ million sources estimates are provided of the stellar effective temperature,
extinction, reddening, and radius and luminosity; and for a pre-selected list of $14\,000$ minor
planets in the solar system epoch astrometry and photometry are presented.  Finally, \gdr{2} also
represents a new materialisation of the celestial reference frame in the optical, the {\gaiacrftwo},
which is the first optical reference frame based solely on extragalactic sources. There are notable
changes in the photometric system and the catalogue source list with respect to \gdr{1}, and we
stress the need to consider the two data releases as independent.}
{\gdr{2} represents a major achievement for the {\gaia} mission, delivering on the long standing
promise to provide parallaxes and proper motions for over 1 billion stars, and representing a first
step in the availability of complementary radial velocity and source astrophysical information for a
sample of stars in the {\gaia} survey which covers a very substantial fraction of the volume of our
galaxy.}

\keywords{ catalogs -
astrometry -
techniques: radial velocities -
stars: fundamental parameters -
stars: variables: general -
minor planets, asteroids: general}

\maketitle

\titlerunning{{\gaia} Data Release 2: Summary} 
\authorrunning{{\gaia} Collaboration}

%-------------------------------------------------------------------
%
% Introduction
%
%-------------------------------------------------------------------

\section{Introduction}
\label{sec:intro}

We present the second intermediate {\gaia} data release ({\gaia} Data Release 2, \gdr{2}), which is
based on the data collected during the first 22 months of the nominal mission lifetime
\citep[scientific data collection started in July 2014 and nominally lasts 60 months,
see][]{2016A&A...595A...1G}. \gdr{2} represents the planned major advance with respect to the first
intermediate {\gaia} data release \citep[\gdr{1},][]{2016A&A...595A...2G}, making the leap to a
high-precision parallax and proper motion catalogue for over 1 billion sources, supplemented by
precise and homogeneous multi-band all-sky photometry and a large radial velocity survey at the
bright ($G\lesssim13$) end. The availability of precise fundamental astrophysical information
required to map and understand the Milky Way is thus expanded to a very substantial fraction of the
volume of our galaxy, well beyond the immediate solar neighbourhood. The data diversity of \gdr{2}
is also significantly enhanced with respect to \gdr{1} through the availability of astrophysical
parameters for a large sample of stars, the significant increase in the number and types of variable
stars and their light curves, and the addition for the first time of solar system astrometry and
photometry.

This paper is structured as follows. In \secref{sec:dataprocessing} we provide a short overview of
the improvements and additions to the data processing that led to the production of \gdr{2}. We
summarise the contents of the second data release in \secref{sec:gdrtwosummary} and illustrate the
quality of this release through all-sky maps of source counts and colours in
\secref{sec:sciencedemos}.  In \secref{sec:dr2vsdr1} we discuss the major differences between
\gdr{2} and \gdr{1}, in particular pointing out the evolution of the source list and the need to
always qualify {\gaia} source identifiers with the data release they refer to. The two releases
should be treated as entirely independent catalogues. The known limitations of the second {\gaia}
data release are presented in \secref{sec:gdrlimitations} and additional guidance on the use of the
data is provided in \secref{sec:guidance}. In \secref{sec:access} we provide updates to the {\gaia}
data access facilities and documentation available to the astronomical community. We conclude with a
look ahead at the next release in \secref{sec:conclusions}. Throughout the paper we make reference
to other DPAC papers that provide more details on the data processing and validation for \gdr{2}.
All these papers (together with the present article) can be found in the Astronomy \& Astrophysics
Special edition on \gdr{2}.

%-------------------------------------------------------------------
%
% Data processing for Gaia DR2
%
%-------------------------------------------------------------------

\section{Data processing for \gdr{2}}
\label{sec:dataprocessing}

To provide the context for the description of the data release contents in the next section, we provide
here a summary of the input measurements used and the main additions and improvements implemented in
the data processing for \gdr{2}. We recall that {\gaia} measurements are collected with three
instruments. The astrometric instrument collects images in {\gaia}'s white-light $G$-band
(330--1050~nm); the Blue (BP) and Red (RP) prism photometers collect low resolution
spectrophotometric measurements of source spectral energy distributions over the wavelength ranges
330--680 nm and 630--1050 nm, respectively; and the radial velocity spectrometer (RVS) collects
medium resolution ($R\sim11\,700$) spectra over the wavelength range 845--872 nm centred on the
Calcium triplet region. For more details on the {\gaia} instruments and measurements we refer to
\cite{2016A&A...595A...1G}. The RVS, from which results are presented in \gdr{2} for the first
time, is described in detail in \cite{DR2-DPACP-46}. An important part of the pre-processing for all
{\gaia} instruments is to remove the effect of non-uniformity of the CCD bias levels, which is
essential for achieving the ultimate image location and radial velocity determination performance.
The details of this process are described in \cite{DR2-DPACP-29}.

The timing of events on board {\gaia}, including the data collection, is given in terms of the
on board mission time line (OBMT) which is generated by the {\gaia} on board clock. By convention
OBMT is expressed in units of 6 h ($21\,600$~s) spacecraft revolutions \citep{2016A&A...595A...1G}.
The approximate relation between OBMT (in revolutions) and the barycentric coordinate time (TCB, in
Julian years) at {\gaia} is
\begin{equation}
  \text{TCB} \simeq \text{J}2015.0 + (\text{OBMT} - 1717.6256~\text{rev})/(1461~\text{rev yr}^{-1})\,.
\end{equation}
The 22 month time interval covered by the observations used for \gdr{2} starts at OBMT
1078.3795~rev = J2014.5624599~TCB (approximately 2014~July~25, 10:30:00~UTC), and ends at OBMT
3750.5602~rev = J2016.3914678~TCB (approximately 2016~May~23, 11:35:00~UTC). As discussed in
\cite{2016A&A...595A...2G} this time interval contains gaps caused by both spacecraft events and by
on-ground data processing problems. This leads to gaps in the data collection or stretches of time
over which the input data cannot be used. Which data are considered unusable varies across the
{\gaia} data processing systems (astrometry, photometry, etc) and as a consequence the effective
amount of input data used differs from one system to the other. We refer to the specific data
processing papers (listed below) for the details.

A broad overview of the data processing for {\gaia} is given in \cite{2016A&A...595A...1G} while the
simplified processing for \gdr{1} is summarised in \cite{2016A&A...595A...2G}, in particular in
their figure 10. With respect to \gdr{1} the following major improvements were implemented in the
astrometric processing \citep[for details, see][]{DR2-DPACP-51}:
\begin{itemize}
  \item Creation of the source list: this process \citep[also known as
    cross-matching;][]{2016A&A...595A...3F} provides the link between the individual {\gaia}
    detections and the entries (`sources') in the {\gaia} working catalogue. For \gdr{1} the
    detections were matched to the nearest source, using a match radius of 1.5~arcsec, and new
    sources were created when no match was found. Spurious detections and limitations of the initial
    source list resulted in many spurious sources but also the loss in \gdr{1} of many real sources,
    including high proper motion stars. For \gdr{2} the source list was created essentially from
    scratch, based directly on the detections and using a cluster analysis algorithm that takes into
    account a possible linear motion of the source. The source list for \gdr{2} is therefore much
    cleaner and of higher angular resolution (Sect.~\ref{sec:sourcelist}), resulting in improved
    astrometry.
  \item Attitude modelling: in the astrometric solution, the pointing of the instrument is modelled
    as a function of time using splines. However, these cannot represent rapid variations caused by
    the active attitude control, micro-clanks (microscopic structural changes in the spacecraft),
    and micrometeoroid hits.  In \gdr{1} the accuracy of the attitude determination was limited by
    such effects. For \gdr{2} the rapid variations are determined and subtracted by a dedicated
    process, using rate measurements from successive CCD observations of bright sources.
  \item Calibration modelling: optical aberrations in the telescopes and the wavelength-dependent
    diffraction create colour-dependent shifts of the stellar images (chromaticity). This will
    eventually be handled in the pre-processing of the raw data, by fitting colour-dependent PSFs or
    LSFs to the CCD samples.  This procedure will only be in place for the next release, and the
    effect was completely ignored for \gdr{1}. In the current astrometric solution chromaticity is
    handled by the introduction of colour-dependent terms in the geometric calibration model.
  \item Global modelling: the basic-angle variations are more accurately modelled thanks to an
    improved processing of the on-board measurements (using the Basic Angle Monitor) and the
    introduction of global corrections to these measurements as additional unknowns in the
    astrometric solution. This has been especially important for reducing large-scale systematics in
    the parallaxes.
  \item Celestial reference frame: establishing a link to the extragalactic reference frame was
    complicated and indirect in \gdr{1}, which relied on the {\hip} and {\tyctwo}
    catalogues for the determination of proper motions. By contrast, \gdr{2} contains the positions
    and proper motions for about half a million identified quasars, which directly define a very
    accurate celestial reference frame (\gaiacrftwo), as described in \cite{DR2-DPACP-30}.
\end{itemize}
The various improvements in the astrometric models have reduced the RMS residual of typical
observations of bright stars ($G\lesssim 13$) from about 0.67~mas in \gdr{1} to 0.2--0.3~mas in
\gdr{2}.

Additional improvements in the data processing for \gdr{2} as well as the introduction of new
elements facilitated the much expanded variety of data published in this second release. Although
the photometric processing pipeline did treat the data from {\gaia}'s BP and RP photometers from the
start of the mission operations, it was decided not to publish the results in \gdr{1}
\citep{2017A&A...600A..51E} because of the still preliminary nature of the calibrations of these
instruments. The processing for \gdr{2} features enhancements in the photometric calibrations,
including of the BP and RP prism spectra. The integrated light from these spectra is published in
this release as the fluxes in the {\gbp} and {\grp} passbands. In addition the photometric passbands
for $G$, {\gbp}, and {\grp} are published, both the versions used in the data processing and the
revised versions (based on a deeper analysis involving the BP/RP spectra of standard stars). The
photometric data processing and results validation for \gdr{2} are described in
\cite{DR2-DPACP-40} and \cite{DR2-DPACP-44}. 

The processing of RVS data was also in place from the start of mission operations but during the
operations up to \gdr{1} the adaptations necessary to the RVS pipeline to deal with the effects of
the excess stray light on board {\gaia} prevented the publication of results. Hence \gdr{2}
features the first RVS results in the form of median radial velocities. The details of the RVS data
processing and results validation are provided in \cite{DR2-DPACP-47}, \cite{DR2-DPACP-54}, and
\cite{DR2-DPACP-48}. 

Epoch astrometry was determined for a list of $14\,000$ pre-selected small solar system bodies
(henceforth referred to as Solar System Objects or SSOs). The data processing and validation for the
\gdr{2} SSO data are described in \cite{DR2-DPACP-32}.

Astrophysical parameters (\teff, \ag, \ebpminrp, radius and luminosity) were determined for between
$77$ and $161$ million stars from the {\gaia} broad-band photometry and parallaxes alone (no
non-{\gaia} data was used). The details of the astrophysical parameter estimation and the validation
of the results are described in \cite{DR2-DPACP-43}. 

Practically all sources present in \gdr{2} were analysed for apparent brightness variations,
resulting in a catalogue of about $0.5$ million stars securely identified as variables and for which
light curves and statistical information on the photometric time series are provided. The
variability processing is described in \cite{DR2-DPACP-49}.

Finally, an overall validation of the \gdr{2} catalogue is described in \cite{DR2-DPACP-39}, which, as
outlined in \cite{2016A&A...595A...1G}, involves an extensive scientific validation of the combined
data presented in this data release.

A number of important shortcomings remain in the data processing, leading to limitations in
\gdr{2} which require taking some care when using the data. In \secref{sec:gdrlimitations} we
summarise the known limitations of the present {\gaia} data release and point out, where relevant,
the causes. \secrefalt{sec:guidance} provides additional guidance on the use of \gdr{2} results.
The reader is strongly encouraged to read the papers listed above and the online
documentation\footnote{\url{http://gea.esac.esa.int/archive/documentation/GDR2/index.html}} to
understand the limitations in detail.

\begin{table}[t]
  \caption{The number of sources of a given type or the number for which a given data product is
    available in \gdr{2}.}
  \label{tab:gdr2stats}
  \centering
  \begin{tabular}{lr}
    \hline\hline
    \noalign{\smallskip}
    Data product or source type & Number of sources \\
    \noalign{\smallskip}
    \hline
    \noalign{\smallskip}
    Total & \gdrtwototnum \\
    \noalign{\smallskip}
    5-parameter astrometry & \gdrtwofivepnum \\
    2-parameter astrometry & \gdrtwotwopnum \\
    ICRF3 prototype sources & \gdrtwoicrfthree \\
    {\gaiacrftwo} sources & \gdrtwogaiacrftwo \\
    \noalign{\smallskip}
    $G$-band & \gdrtwogbandnum \\
    \gbp-band & \gdrtwobpbandnum \\
    \grp-band & \gdrtworpbandnum \\
    \noalign{\smallskip}
    Radial velocity & \gdrtwovradnum \\
    \noalign{\smallskip}
    Classified as variable & \gdrtwovarnum \\
    Variable type estimated & \gdrtwovartypednum \\
    Detailed characterisation of light curve & \gdrtwovarsostotnum \\
    \noalign{\smallskip}
    Effective temperature {\teff} & \gdrtwoteffnum \\
    Extinction {\ag} & \gdrtwoagnum \\
    Colour excess {\ebpminrp} & \gdrtwoebpminrpnum \\
    Radius & \gdrtworadiusnum \\
    Luminosity & \gdrtwolumnum \\
    \noalign{\smallskip}
    SSO epoch astrometry and photometry & \gdrtwossonum \\
    \noalign{\smallskip}
    \hline
  \end{tabular}
\end{table}

\begin{table}[t]
  \caption{The distribution of the \gdr{2} sources in $G$-band magnitude. The distribution
percentiles are shown for all sources and for those with a 5-parameter and 2-parameter astrometric
solution, respectively.}
  \label{tab:gdr2magperc}
  \centering
  \begin{tabular}{lrrr}
    \hline\hline
    \noalign{\smallskip}
    \multicolumn{4}{c}{Magnitude distribution percentiles ($G$)} \\
    Percentile & All & 5-parameter & 2-parameter \\
    \noalign{\smallskip}
    \hline
    \noalign{\smallskip}
    $0.135$\% & $11.6$ & $11.4$ & $15.3$ \\
    $2.275$\% & $15.0$ & $14.7$ & $18.5$ \\
    $15.866$\% & $17.8$ & $17.4$ & $19.8$ \\
    $50$\% & $19.6$ & $19.3$ & $20.6$ \\
    $84.134$\% & $20.6$ & $20.3$ & $21.0$ \\
    $97.725$\% & $21.1$ & $20.8$ & $21.2$ \\
    $99.865$\% & $21.3$ & $20.9$ & $21.4$ \\[3pt]
    \noalign{\smallskip}
    \hline
  \end{tabular}
\end{table}

%-------------------------------------------------------------------
%
% Overview of the contents of Gaia DR2
%
%-------------------------------------------------------------------

\section{Overview of the contents of \gdr{2}}
\label{sec:gdrtwosummary}

\begin{figure*}[t]
  \sidecaption
  \includegraphics[width=12cm]{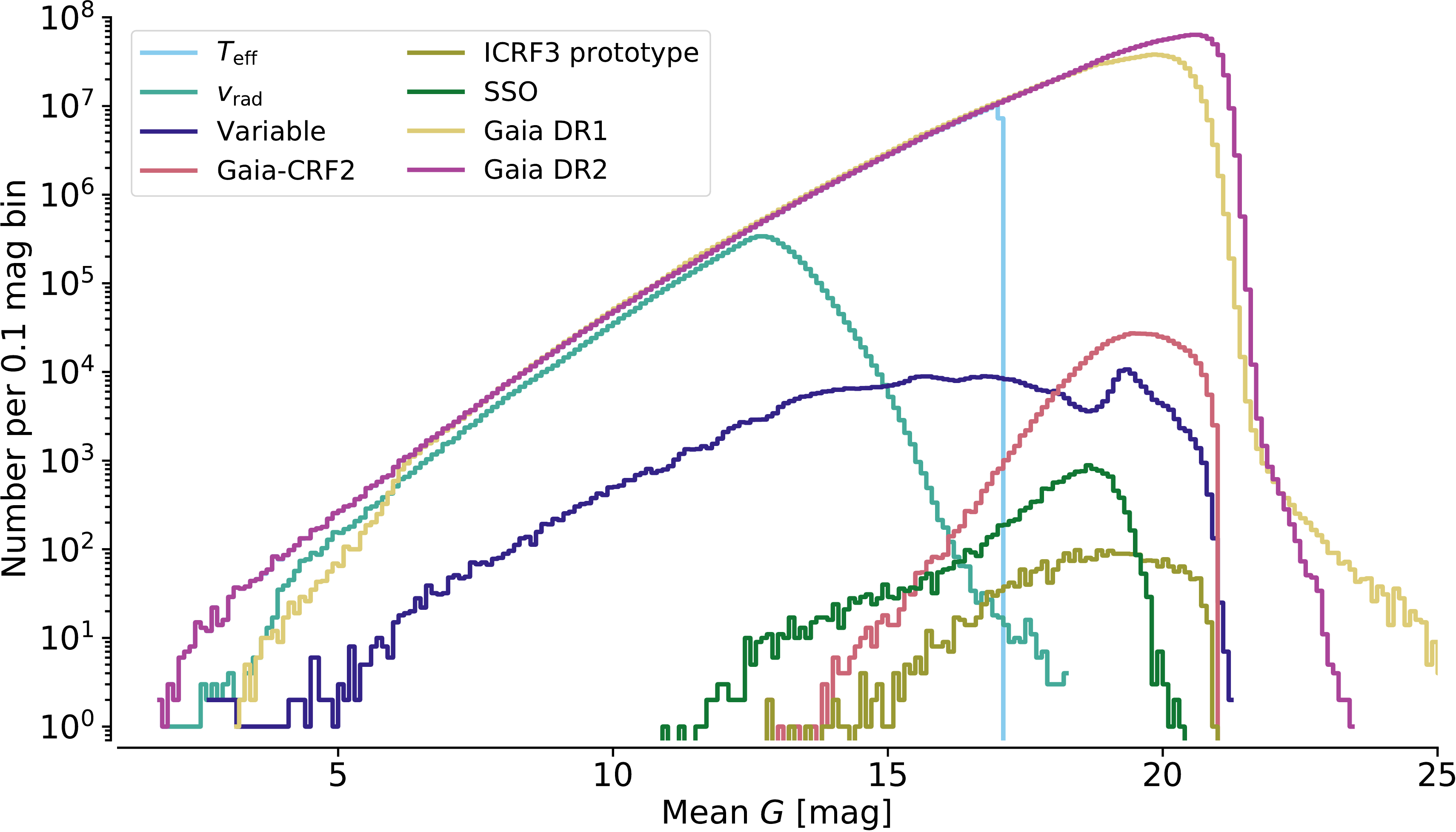}
  \caption{Distribution of the mean values of $G$ for all \gdr{2} sources shown as histograms with
    $0.1$~mag wide bins. The distribution of the \gdr{1} sources is included for comparison and
    illustrates the improved photometry at the faint end and the improved completeness at the bright
    end. The other histograms are for the main \gdr{2} components as indicated in the legend. See
  text for further explanations on the characteristics of the histograms.\label{fig:gmaghistos}}
\end{figure*}

\begin{figure}[t]
  \resizebox{\hsize}{!}{\includegraphics{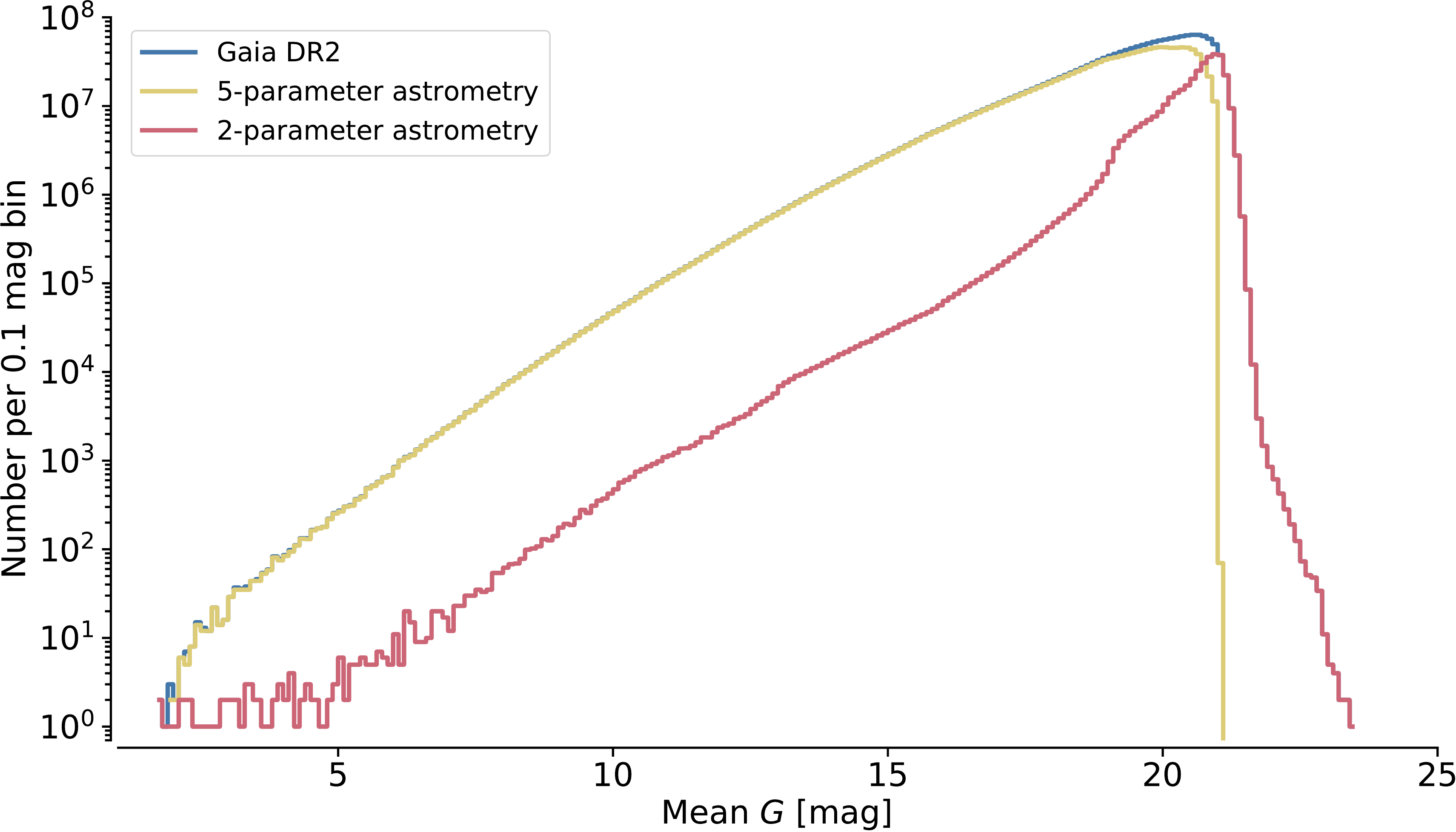}}
  \caption{Distribution of the mean values of $G$ for the sources with a full astrometric solution
    in \gdr{2} (`5-parameter') and for the sources for which only the celestial position is listed
    (`2-parameter') compared to the overall magnitude distribution for \gdr{2}.
  \label{fig:gmaghistoastrometry}}
\end{figure}

\gdr{2} contains astrometry, broad-band photometry, radial velocities, variable star
classifications as well as the characterisation of the corresponding light curves, and astrophysical
parameter estimates for a total of {\gdrtwototnum} sources. In addition the epoch astrometry and
photometry for {\gdrtwossonum} solar system objects are listed. Basic statistics on the source
numbers and the overall distribution in $G$ can be found in \tabref{tab:gdr2stats} and
\tabref{tab:gdr2magperc}, where it should be noted that 4 per cent of the sources are fainter than
$G=21$. The overall quality of \gdr{2} results in terms of the typically
achieved uncertainties is summarised in \tabref{tab:qualitystats}. The contents of the main
components of the release, of which the magnitude distributions are shown in \figsref{fig:gmaghistos}
and \ref{fig:gmaghistoastrometry}, are summarised in the following paragraphs. We defer the
discussion on the known limitations of \gdr{2} to \secref{sec:gdrlimitations}.

\subsection{Astrometric data set} 

The astrometric data set consists of two subsets: for {\gdrtwofivepnum} sources the full
five-parameter astrometric solution is provided (`5-parameter' in \tabref{tab:gdr2stats}), hence
including celestial position, parallax, and proper motion. For the remaining {\gdrtwotwopnum}
sources (`2-parameter' in \tabref{tab:gdr2stats}) only the celestial positions $(\alpha,\delta)$ are
reported. \figrefalt{fig:gmaghistoastrometry} shows the distribution in $G$ for the 5-parameter and
2-parameter sources compared to the overall magnitude distribution. The 2-parameter sources are
typically faint (with about half those sources at $G>20.6$, see \tabref{tab:gdr2magperc}), have very
few observations, or very poorly fit the five-parameter astrometric model. All sources fainter than
$G=21$ have only positions in \gdr{2}. We refer to \cite{DR2-DPACP-51} for the detailed criteria
used during the data processing to decide which type of solution should be adopted.

For a 2-parameter source the position was computed using a special fall-back solution. Rather than
ignoring the parallax and proper motion of the source (i.e.\ assuming that they are strictly zero),
the fall-back solution estimates all five parameters but applies a prior that effectively constrains
the parallax and proper motion to realistically small values, depending on the magnitude and
Galactic coordinates of the source \citep{2015A&A...583A..68M}. The resulting position is usually
more precise, and its uncertainty more realistic (larger), than if only the position had been
solved for. The parallax and proper motion of the fall-back solution may however be strongly biased,
which is why they are not published.

The reference epoch for all (5- and 2-parameter) sources is J2015.5 (TCB).  This epoch, close to the
mid-time of the observations included in \gdr{2}, was chosen to minimise correlations between the
position and proper motion parameters. This epoch is 0.5~year later than the reference epoch for
\gdr{1}, which must be taken into account when comparing the positions between the two releases.

As for \gdr{1} all sources were treated as single stars when solving for the astrometric parameters.
For a binary the parameters may thus refer to either component, or to the photocentre of the system,
and the proper motion represents the mean motion of the component, or photocentre, over the
1.75~years of data included in the solution. Depending on the orbital motion, this could be
significantly different from the proper motion of the same object in \gdr{1} (see
Sect.~\ref{sec:dr2vsdr1}).

The positions and proper motions are given in the second realisation of the {\gaia} celestial
reference frame (\gaiacrftwo) which at the faint end ($G\sim19$) is aligned with the International
Celestial Reference Frame (ICRF) to about $0.02$~mas RMS at epoch J2015.5 (TCB), and non-rotating with
respect to the ICRF to within $0.02$~\masyr RMS. At the bright end ($G<12$) the alignment can only be
confirmed to be better than $0.3$~mas while the bright reference frame is non-rotating to within
$0.15$~\masyr. For details we refer to \cite{DR2-DPACP-51}. The {\gaiacrftwo} is materialised by
{\gdrtwogaiacrftwo} QSOs and aligned to the forthcoming version 3 of the ICRF through a subset of
{\gdrtwoicrfthree} QSOs. It represents the first ever optical reference frame constructed on the
basis of extragalactic sources only. The construction and properties of the {\gaiacrftwo} as well as
the comparison to the ICRF3 prototype are described in \cite{DR2-DPACP-30}.

\begin{table*}[ht]
  \caption{Basic performance statistics for \gdr{2}. The astrometric uncertainties as well as the
    \gaia-CRF2 alignment and rotation limits refer to epoch J2015.5 TCB. The uncertainties on the
    photometry refer to the mean magnitudes listed in the main \gdr{2}
  catalogue.\label{tab:qualitystats}}
  \centering
  \begin{tabular}{lr}   
    \hline\hline        
    \noalign{\smallskip}
    Data product or source type                                               & Typical uncertainty\\
    \noalign{\smallskip}
    \hline
    \noalign{\smallskip}
    Five-parameter astrometry (position \& parallax)                          & $0.02$--$0.04$~mas at $G<15$\\
                                                                              & $0.1$~mas at $G=17$\\
                                                                              & $0.7$~mas at $G=20$\\
                                                                              & $2$~mas at
                                                                              $G=21$\\
    \noalign{\smallskip}
    Five-parameter astrometry (proper motion)                                 & $0.07$~mas~yr$^{-1}$ at $G<15$\\
                                                                              & $0.2$~~mas~yr$^{-1}$ at $G=17$\\
                                                                              & $1.2$~mas~yr$^{-1}$ at $G=20$\\
                                                                              & $3$~mas~yr$^{-1}$ at $G=21$\\
    \noalign{\smallskip}
    Two-parameter astrometry (position only)                                  & $1$--$4$~mas\\
    \noalign{\smallskip}
    Systematic astrometric errors (averaged over the sky)                     & $<0.1$~mas\\
    \noalign{\smallskip}
    \gaia-CRF2 alignment with ICRF                                            & $0.02$~mas at $G=19$\\
    \gaia-CRF2 rotation with respect to ICRF                                  & $<0.02$~mas~yr$^{-1}$ at $G=19$\\
    \gaia-CRF2 alignment with ICRF                                            & $0.3$~mas at $G<12$\\
    \gaia-CRF2 rotation with respect to ICRF                                  & $<0.15$~mas~yr$^{-1}$ at $G<12$\\
    \noalign{\smallskip}
    Mean $G$-band photometry                                                  & $0.3$~mmag at $G<13$\\
    & $2$~mmag at $G=17$\\
    & $10$~mmag at $G=20$\\
    \noalign{\smallskip}
    Mean \gbp- and \grp-band photometry                                       & $2$~mmag at $G<13$\\
                                                                              & $10$~mmag at $G=17$\\
    & $200$~mmag at $G=20$\\
    \noalign{\smallskip}
    Median radial velocity over 22 months                                     & $0.3$~km~s$^{-1}$ at $G_{\rm RVS} < 8$\\
                                                                              & $0.6$~km~s$^{-1}$ at $G_{\rm RVS} = 10$\\
    & $1.8$~km~s$^{-1}$ at $G_{\rm RVS} = 11.75$\\
    \noalign{\smallskip}
    Systematic radial velocity errors                                         & $<0.1$~km~s$^{-1}$ at $G_{\rm RVS} < 9$\\
    & $0.5$~km~s$^{-1}$ at $G_{\rm RVS} = 11.75$\\
    \noalign{\smallskip}
    Effective temperature {\ensuremath{T_\mathrm{eff}}}                       & 324~K\\
    Extinction {\ensuremath{A_G}}                                             & 0.46~mag\\
    Colour excess {\ensuremath{E(G_\mathrm{BP}-G_\mathrm{RP})}}               & 0.23~mag\\
    Radius                                                                    & 10\%\\
    Luminosity                                                                & 15\%\\
    \noalign{\smallskip}
    Solar system object epoch astrometry                                      & 1~mas (in scan direction)\\
    \hline
  \end{tabular}
\end{table*}

\subsection{Photometric data set}
\label{sec:photdata}

The photometric data set contains the broad band photometry in the $G$, {\gbp}, and {\grp} bands,
thus providing the major new element of colour information for \gdr{2} sources. The mean value of
the $G$-band fluxes is reported for all sources while for about 80 per cent of the sources the mean
values of the {\gbp} and {\grp} fluxes are provided (for a small fraction of these sources only
the {\grp} value is reported). The photometric data processing considered three types of sources,
`Gold', `Silver', and `Bronze', which represent decreasing quality levels of the photometric
calibration achieved, where in the case of the Bronze sources no colour information is available. 
The photometric nature of each source is indicated in the released catalogue by a numeric field
(\texttt{phot\_proc\_mode}) assuming values 0, 1 and 2 for gold, silver, and bronze sources
respectively. At the bright end the photometric uncertainties are dominated by calibration effects
which are estimated to contribute $2$, $5$, and $3$ mmag RMS per CCD observation, respectively for
$G$, {\gbp}, and {\grp} \citep{DR2-DPACP-40}. For details on the photometric processing and the
validation of the results we refer to \cite{DR2-DPACP-44} and \cite{DR2-DPACP-40}.

The broad-band colour information suffers from strong systematic effects at the faint end of the
survey ($G\gtrsim19$), in crowded regions, and near bright stars. In these cases the photometric
measurements from the blue and red photometers suffer from an insufficiently accurate background
estimation and from the lack of specific treatment of the prism spectra in crowded regions, where
the overlapping of images of nearby sources is not yet accounted for. This leads to measured fluxes
that are inconsistent between the $G$ and the {\gbp} and {\grp} bands in the sense that the sum of
the flux values in the latter two bands may be significantly larger than that in $G$ (whereas it is
expected that for normal spectral energy distributions the sum of fluxes in $\gbp$ and {\grp} should
be comparable to that in $G$). A quantitative indication of this effect is included in \gdr{2} in
the form of the `flux excess factor' (the \texttt{phot\_bp\_rp\_excess\_factor} field in the
data archive).

The distribution of the astrometric and photometric data sets in $G$ is shown in purple in
\figref{fig:gmaghistos}, where for comparison the distribution for \gdr{1} is also shown in yellow.
Note the improved completeness at the bright end of the survey and the improved photometry (less
extremely faint sources) and completeness at the faint end. The distribution of the {\gaiacrftwo}
sources (pink-red line) shows a sharp drop at $G=21$ which is because only QSOs at $G<21$ were used
for the construction of the reference frame.

\subsection{Radial velocity data set}
\label{sec:rvsdata}

The radial velocity data set contains the median radial velocities, averaged over the 22 month time
span of the observations, for {\gdrtwovradnum} sources which are nominally brighter than 12th
magnitude in the {\grvs} photometric band. For the selection of sources to process, the provisional
{\grvs} magnitude as listed in the Initial Gaia Source List \citep{2014AA...570A..87S} was used.
The actual magnitudes in the {\grvs} band differ from these provisional values, meaning that the
magnitude limit in {\grvs} is not sharply defined. In practice the sources for which a median radial
velocity is listed mostly have magnitudes brighter than 13 in $G$ (see light green line in
\figref{fig:gmaghistos}). The signal to noise ratio of the RVS spectra depends primarily on {\grvs},
which is not listed in \gdr{2}. It was decided not to publish the {\grvs} magnitude in \gdr{2}
because the processing of RVS data was focused on the production of the radial velocities, and the
calibrations necessary for the estimation of the flux in the RVS passband (background light
corrections and the knowledge of the PSF in the direction perpendicular to Gaia's scanning
direction) were only preliminary. As a result the {\grvs} magnitudes were of insufficient quality
for publication in \gdr{2} \citep{DR2-DPACP-47}.  The value of {\grvs} as determined during the data
processing was however used to filter out stars considered too faint ($\grvs>14$) for inclusion in
the radial velocity data set. For convenience we provide here a relation which allows to predict the
value of {\grvs} from the $(G-\grp)$ colour.
\begin{multline}
  \grvs-\grp = 0.042319 - 0.65124(G-\grp) + 1.0215(G-\grp)^2\\
  -1.3947 (G-\grp)^3 + 0.53768(G-\grp)^4 \\
  \text{to within }0.086\text{ mag RMS for }0.1<(G-\grp)<1.4\,,
\end{multline}
and
\begin{multline}
  \grvs-\grp = 132.32 - 377.28(G-\grp) + 402.32(G-\grp)^2\\
  -190.97(G-\grp)^3 + 34.026(G-\grp)^4\\
  \text{to within }0.088\text{ mag RMS for }1.4\leq(G-\grp)<1.7\,.
  \label{eq:rvsfromgmingrp}
\end{multline}
This relation was derived from a sample of stars for which the flux in the RVS band could be
determined to a precision of $0.1$~mag or better.

Radial velocities are only reported for stars with effective temperatures in the range
$3550$--$6900$~K (where these temperatures refer to the spectral template used in the processing,
not to the {\teff} values reported as part of the astrophysical parameter data set). The
uncertainties of the radial velocities are summarised in \tabref{tab:qualitystats}. At the faint end
the uncertainties show a dependency on stellar effective temperature, where the values are
approximately $1.4$~\kms\ and $3.6$~\kms\ at $\grvs=11.75$ for stars with $\teff\sim5500$~K and
$\teff\sim6500$~K, respectively. The distribution over $G$ of the sources with radial velocities
shown in \figref{fig:gmaghistos} in light green reflects the fact that over the range $4<G<12$ the
completeness of the radial velocity data set with respect to the \gdr{2} data set varies from 60 to
80 per cent \citep{DR2-DPACP-54}. At the faint end ($G>13$) the shape of the distribution is
determined by the selection of stars for which radial velocities were derived (using the provisional
value of \grvs) and the large differences between $G$ and $\grvs$ that can occur depending on the
effective temperature of the stars. For the details on the radial velocity data processing and the
properties and validation of the resulting radial velocity catalogue we refer to \cite{DR2-DPACP-47}
and \cite{DR2-DPACP-54}. The set of standard stars that was used to define the zeropoint of the RVS
radial velocities is described in \cite{DR2-DPACP-48}.

\subsection{Variability data set} 
\label{sec:varidata}

The variability data set consists of {\gdrtwovarnum} sources that are securely identified as
variable (based on at least two transits of the sources across the fields of view of the two {\gaia}
telescopes) and for which the photometric time series and corresponding statistics are provided.
This number still represents only a small subset of the total amount of variables expected in the
{\gaia} survey and subsequent data releases will contain increasing numbers of variable sources. Of
the sources identified as variable {\gdrtwovartypednum} were classified into one of nine variable
types by a supervised light curve classifier. The types listed in the \gdr{2} are: RR Lyrae
(anomalous RRd, RRd, RRab, RRc); long period variables (Mira type and Semi-Regulars); Cepheids
(anomalous Cepheids, classical Cepheids, type-II Cepheids); $\delta$~Scuti and SX~Phoenicis stars. A
second subset of {\gdrtwovarsostotnum} variable stars (largely overlapping with the variability type
subset) was analysed in detail when at least 12 points were available for the light curve. These
so-called `specific object studies' (SOS) were carried out for variables of the type Cepheid and RR
Lyrae, long period variables, short time scale variables (with brightness variations on time scales
of one day or less), and rotational modulation variables.

\figrefalt{fig:gmaghistos} shows in dark blue the distribution over $G$ of the sources identified as
variable. The mean $G$ value as determined in the photometric data processing (used in
\figref{fig:gmaghistos}) may differ from the mean magnitude determined from the photometric time
series where the variable nature of the source is properly accounted for. Hence the distribution in
\figref{fig:gmaghistos} should be taken as illustrative only. For full details on the variable star
processing and results validation we refer to \cite{DR2-DPACP-49} and references therein.

\subsection{Astrophysical parameter data set} 

The astrophysical parameter data set consists of estimated values of {\teff}, extinction {\ag} and
reddening {\ebpminrp} (both derived from the apparent dimming and reddening of a source), radius,
and luminosity for stars brighter than $G=17$. \tabref{tab:gdr2stats} contains the source counts for
each of these astrophysical parameters. The magnitude distribution shown in \figref{fig:gmaghistos}
in cyan concerns all sources for which {\teff} was estimated and indicates that this parameter is
available for practically all sources at $G<17$.  Values of {\teff} are only reported over the range
$3000$--$10\,000$~K, which reflects the limits of the training data for the algorithm used to
estimate \teff. Estimates of the other astrophysical parameters are published for about 50\% of the
sources for which {\teff} is published. This is caused by the filtering of the pipeline results to
remove parameter estimates for which the input data are too poor or for which the assumptions made
lead to invalid results. The details of the astrophysical parameter processing and the validation of
the results are described in \cite{DR2-DPACP-43}.

\subsection{Solar system objects data set}

The solar system objects data set features epoch astrometry and photometry for a pre-selected list
of {\gdrtwossonum} known minor bodies in the solar system, primarily main belt asteroids. Epoch
astrometry refers to the fact that the measured celestial position for a given SSO is listed for
each instance in time when it passed across the field of view of one of {\gaia}'s telescopes. The
celestial positions at each epoch are given as seen from {\gaia}. These measurements can be used to
determine orbits for the SSOs and the results thereof are described in \cite{DR2-DPACP-32}. For
details on the processing of SSOs we refer to the same paper.  Over the apparent magnitude range
$G\sim12$--$17$ the typical focal plane transit level of uncertainty achieved for the instantaneous
SSO celestial positions is 1~mas in the {\gaia} scanning direction. \figref{fig:gmaghistos} shows in
dark green the magnitude distribution for the SSOs, where it should be noted that the magnitudes as
can be measured by {\gaia} represent instantaneous measurements taken far from opposition. Hence the
magnitude histogram is to be taken as illustrative only.

\begin{figure*}[t]
  \includegraphics[width=\textwidth]{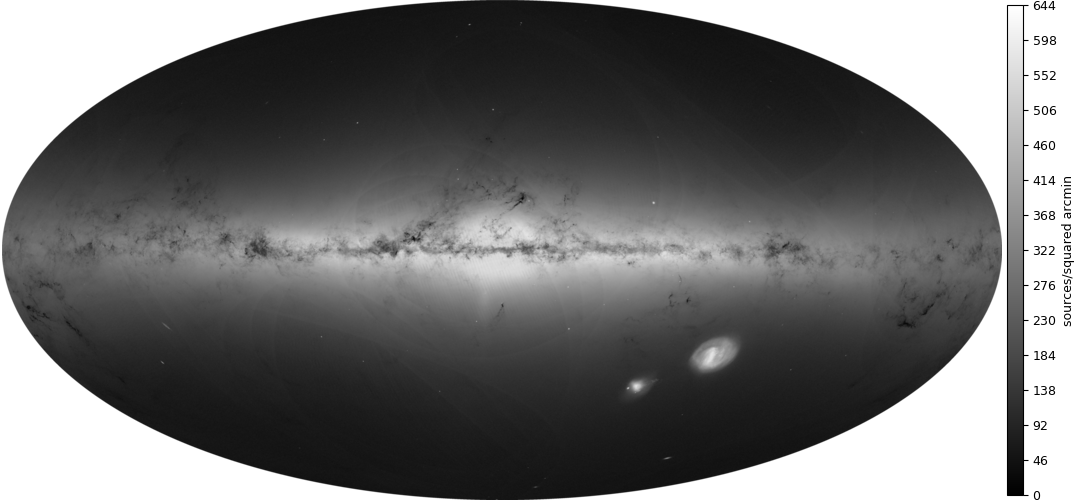}
  \caption{Sky distribution of all \gdr{2} sources in Galactic coordinates. This image and the one
    in \figref{fig:mwcolours} are Hammer projections of the full sky. This projection was chosen in
    order to have the same area per pixel (not strictly true because of pixel discretisation). Each
    pixel is $\sim5.9$ square arcmin. The colour scale is logarithmic and represents the number of
  sources per square arcmin.\label{fig:sourcedensity}}
\end{figure*}

\begin{figure*}[t]
  \includegraphics[width=\textwidth]{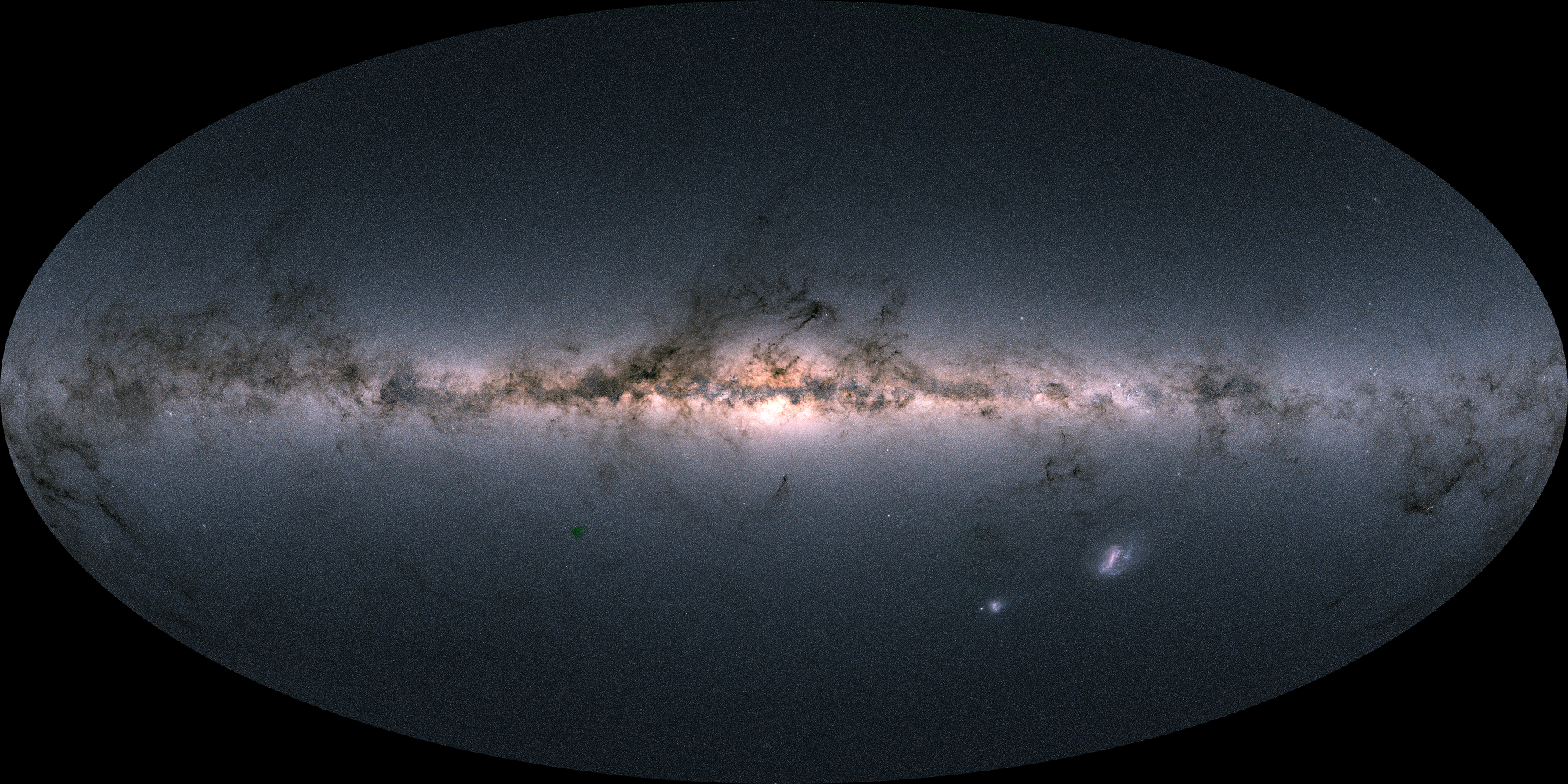}
  \caption{Map of the total flux measured in the {\grp}, $G$, and {\gbp} bands, where the flux in
  these bands is encoded in the red, green, and blue channel, respectively. There is one easily visible
  artefact in this map, a `green' patch to the lower left of the bulge which is a region where
  {\gbp} and {\grp} data are not available for a large number of sources, leading to the greenish
  colour which was used to encode the $G$-band fluxes (which are available for all sources). Such
  artefacts also occur (although not as visible) in the region to the upper left of the Small
  Magellanic Cloud and at high Galactic latitude to the right of the north Galactic pole region. The
  areas where green patches are likely to occur can be identified in Figure 27 in
  \cite{DR2-DPACP-40} which shows the celestial distribution of \gdr{2} sources for which no BP/RP
  photometry is available.\label{fig:mwcolours}}
\end{figure*}

%-------------------------------------------------------------------
%
% Scientific performance and potential of Gaia DR2
%
%-------------------------------------------------------------------

\section{Scientific performance and potential of \gdr{2}}
\label{sec:sciencedemos}

\gdr{2} is accompanied by six papers that provide basic demonstrations of the scientific quality
of the results included in this release. The topics treated by the papers are: 
\begin{itemize}
  \item the reference frame {\gaiacrftwo} \citep{DR2-DPACP-30}; 
  \item orbital fitting of the epoch astrometry for solar system objects \citep{DR2-DPACP-32};
  \item variable stars as seen in the \gdr{2} colour-magnitude diagram \citep{DR2-DPACP-35}, where
    the motion of variables in colour-magnitude space is explored;
  \item the kinematics of the Milky Way disk \citep{DR2-DPACP-33}, illustrating in particular the
    power of having radial velocities available in \gdr{2};
  \item the kinematics of globular clusters, the LMC and SMC, and other dwarf galaxies around the
    Milky Way \citep{DR2-DPACP-34}, showcasing the power of \gdr{2} to study distant samples of
    stars; 
  \item the observational Hertzsprung-Russell diagram is explored in
    \cite{DR2-DPACP-31}.
\end{itemize}
We strongly encourage the reader to consult these papers for a full impression of the enormous
scientific potential of the second {\gaia} data release.

Here we restrict ourselves to illustrating both the improvement in the data quality and the expanded
set of data products through the updated map of the {\gaia} sky.  \figrefalt{fig:sourcedensity}
shows the sky distribution of all the sources present in \gdr{2} in the form of source densities on
a logarithmic scale. When comparing to the map produced from \gdr{1} data
\citep{2016A&A...595A...2G} it is immediately apparent that there is a strong reduction in the
artefacts caused by the combination of source filtering and the {\gaia} scanning law \citep[see][for
a more detailed explanation of these artefacts]{2016A&A...595A...2G}, which is another illustration
of the increased survey completeness of \gdr{2}. Nonetheless there are still source count variations
visible, which clearly are imprints from the scanning law (as executed over the first 22 months of
the mission). For example there are two arcs above and below the $\rho$~Oph clouds that can be
traced all the way down to and below the Galactic plane (these can best be seen in the electronic
version of the figure). Such arcs occur all along the ecliptic plane and are regions on the sky that
were scanned more frequently by Gaia and therefore contain relatively more sources that were
observed often enough for inclusion in the published catalogue.

One newly visible (and real) feature in this map is the Sagittarius dwarf which can be noted as an
excess in star counts in a strip below the bulge region, stretching to the R Corona Australis
region.

\begin{figure}[t]
  \begin{center}
    \includegraphics[width=0.9\columnwidth]{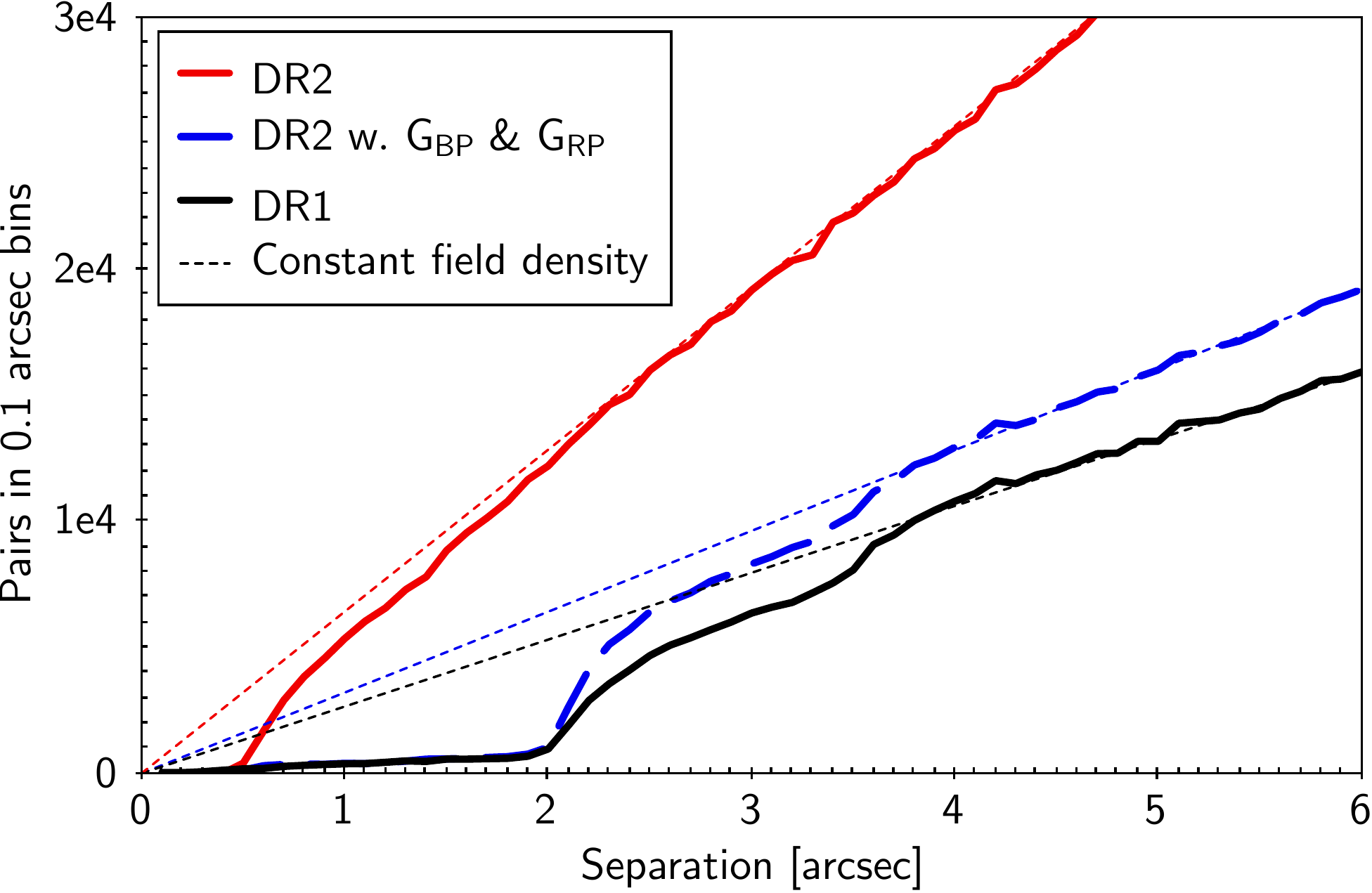}\\[3mm]
    \includegraphics[width=0.9\columnwidth]{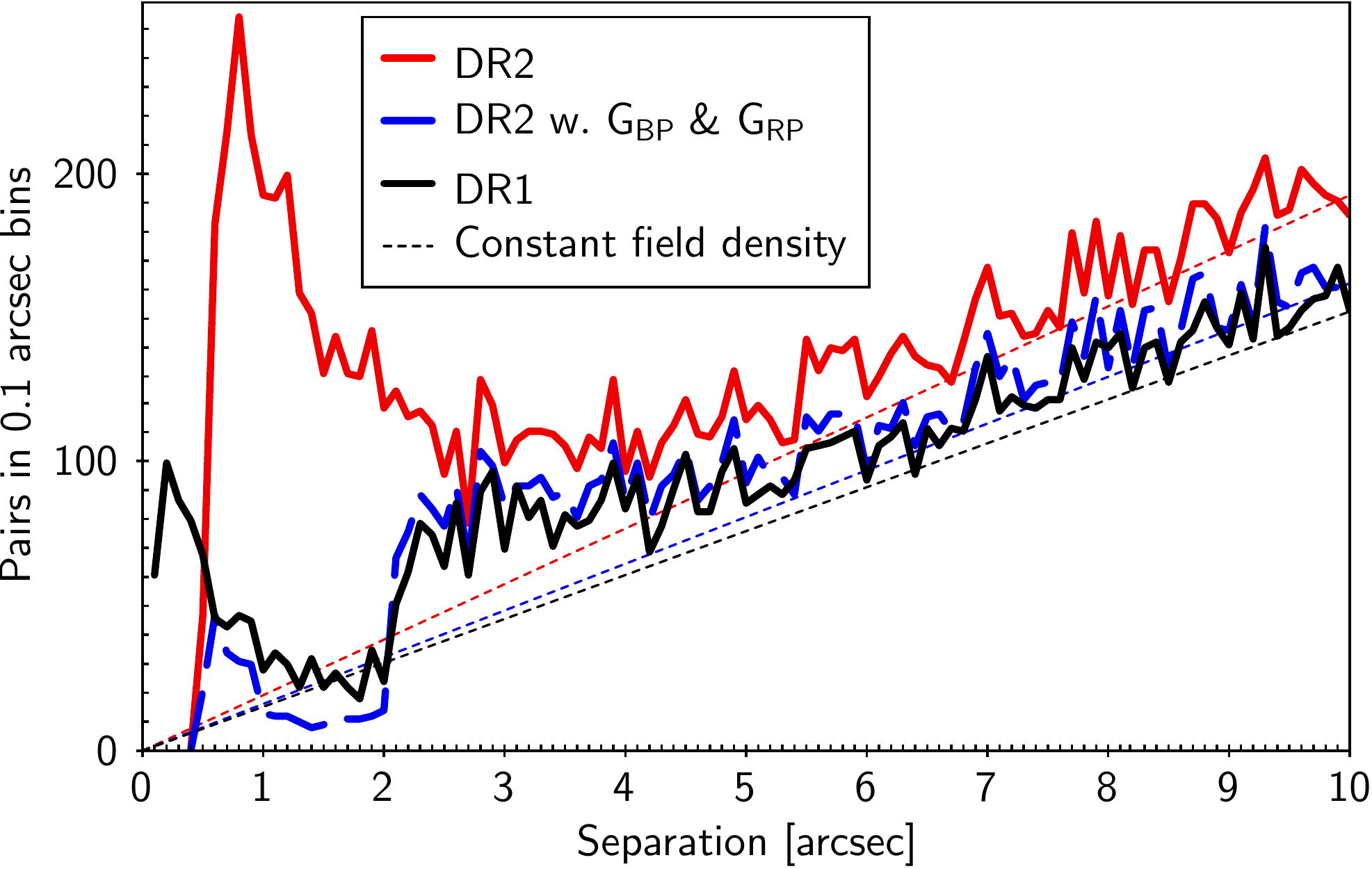}
    \caption{Histograms from \cite{DR2-DPACP-39} of source pair separations in two circular test
      fields for \gdr{2} sources (red lines); \gdr{2} sources with \gbp\ and \grp\ photometry (blue
      lines); and \gdr{1} sources (black lines). {\em Top:} a dense field of radius 0.5\degr\ at
      $(\ell,b) = (-30\degr,-4\degr)$ with 456\,142 sources, {\em Bottom:} a sparse field of radius
      5\degr\ at $(\ell,b) = (-100\degr,-60\degr)$ with 250\,092 sources. The thin, dotted lines
    show the relations for a constant density across the field.\label{fig:pairstat}}
  \end{center}
\end{figure}

\figrefalt{fig:mwcolours} shows a map that combines the integrated fluxes as observed in the {\grp},
$G$, and {\gbp} bands, where the integrated flux map for each of the bands was used to colour code
the image according to a red, green, and blue channel. The map illustrates the availability of
homogeneous all-sky multi-band photometry in \gdr{2} and offers a magnificent view of the Milky
Way in colour. This flux map also reveals numerous open clusters which are not readily visible in
the source count map (while on the other hand many faint source concentrations, such as distant
dwarf galaxies are no longer visible). Complete details on the construction of the images in
\figsref{fig:sourcedensity} and \ref{fig:mwcolours} are provided in \cite{DR2-DPACP-42}.

One aspect of the sky maps shown in \figsref{fig:sourcedensity} and \ref{fig:mwcolours} that is
perhaps not as well appreciated is their effective angular resolution, which given the size of
{\gaia}'s main telescope mirrors \citep[$1.45$~m along the scanning
direction,][]{2016A&A...595A...1G} should be comparable to that of the Hubble Space Telescope.
\cite{2016A&A...595A...2G} and \cite{2017A&A...599A..50A} discuss how the effective angular
resolution of \gdr{1} is limited to about 2--4 arcsec owing to limitations in the data processing.
This has much improved for \gdr{2}.  The gain in angular resolution is illustrated in
\figref{fig:pairstat}. The top panel shows the distribution of source pair distances in a small,
dense field. For \gdr{2} (upper, red curve) source pairs below $0.4$--$0.5$ arcsec are rarely
resolved, but the resolution improves rapidly and above $2.2$ arcsec practically all pairs are
resolved.  For \gdr{1} the fraction of resolved source pairs started to fail at separations of
$3.5$ arcsec, reaching very low values below $2.0$ arcsec. The same, modest resolution is seen for
\gdr{2} if we only consider sources with \gbp\ and \grp\ photometry. The reason is the angular
extent of the prism spectra and the fact that \gdr{1} only includes sources for which the
integrated flux from the BP/RP spectra could be reliably determined. The lower panel shows in the
same way the source pairs in the one hundred times larger, sparse field.  The more remarkable
feature here is the peak of resolved binaries at small separations, which was missed in \gdr{1}. A
similar population must be present in the dense field, where it cannot be discerned because the
field is dominated by distant sources. The figure also demonstrates that the gain in number of
sources from \gdr{1} to \gdr{2} is mainly due to the close source pairs.  Finally,
\figref{fig:pairstat} clearly demonstrates that the effective angular resolution of \gdr{2} quite
significantly exceeds that of all ground-based large-area optical sky surveys.

%-------------------------------------------------------------------
%
% Treat Gaia DR2 as independent from Gaia DR1
%
%-------------------------------------------------------------------

\section{Treat \gdr{2} as independent from \gdr{1}}
\label{sec:dr2vsdr1}

Although \gdr{1} and \gdr{2} are based on observations from the same instruments, the discussion
in the following subsections shows that the two releases should be treated as independent. In
particular the tracing of sources from \gdr{1} to \gdr{2} (should this be needed for a
particular application) must be done with care.

\subsection{\gdr{2} represents a stand-alone astrometric catalogue}

Because the observational time baseline for \gdr{2} is sufficiently long, parallax and proper
motion can be derived from the {\gaia} observations alone. That is, the {\tyc}-{\gaia} Astrometric
Solution \citep[TGAS,][]{2015A&A...574A.115M} as employed for the 2 million brightest stars in
\gdr{1} is no longer needed, and the astrometric results reported in \gdr{2} are based solely on
{\gaia} observations. For the TGAS subset from \gdr{1} there is thus a large difference in the
time baseline for the proper motions ($\sim24$~yr vs $\sim 2$~yr) which means there can be significant
differences between TGAS and \gdr{2} proper motions for binary stars with orbital periods
comparable to 2 years. The TGAS proper motions may be more reliable in such cases. However,
discrepancies can also point to erroneous TGAS proper motions related to a mismatching between
(components of) sources observed by {\gaia} and {\hip} \citep[see][for a discussion of this
issue]{2017ApJ...840L...1M}. In cases where proper motion discrepancies are of interest they should
be carefully investigated before deciding which values to use or concluding that the discrepancy
points to the source not being a single star.

\begin{figure}[t]
  \centering
  \resizebox{\hsize}{!}{\includegraphics{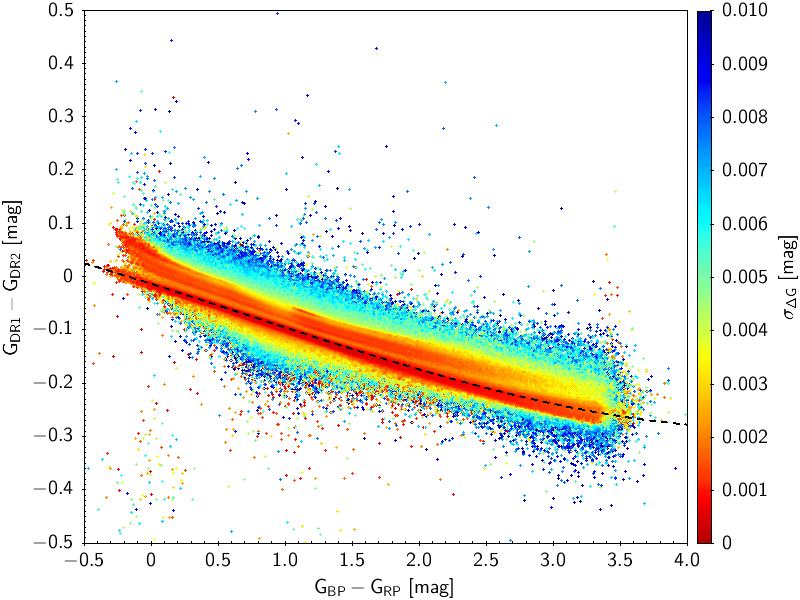}}\\
  \resizebox{\hsize}{!}{\includegraphics{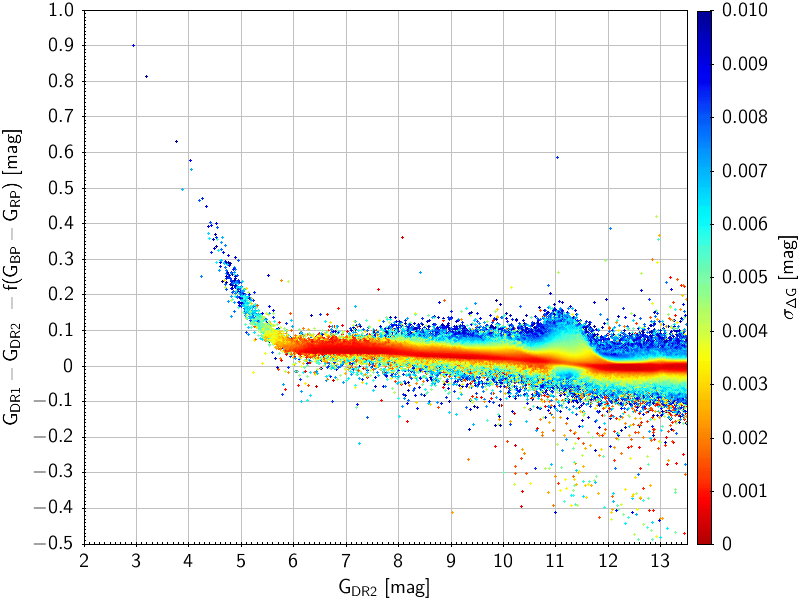}}
  \caption{The top diagram shows the difference in the value of $G$ (with $\Delta G =
    G_\mathrm{DR1}-G_\mathrm{DR2}$) as listed for the same sources in \gdr{1} and \gdr{2} as a
    function of $(\gbp-\grp)$. The source pairs selected from the two releases match in celestial
    position to within $0.25$~arcsec and the formal error on the magnitude differences is less than
    $0.01$. All sources were selected to have a flux excess factor in \gdr{2} of less than $1.6$.
    The dashed line shows a polynomial relation between the difference in $G$ and the colour. The
    colour scale indicates the estimated uncertainty on $\Delta G$. The bottom panel shows the
    relation between $\Delta G$ and $G$ after removing the colour dependency using the polynomial
  relation in the top panel.\label{fig:gdiffs}}
\end{figure}

\subsection{Photometric system evolution}

The photometric data processing for \gdr{2} \citep{DR2-DPACP-44,DR2-DPACP-40} features many improvements
with respect to \gdr{1} and represents a new photometric reduction. In particular more input data
was used and the stretch of data selected for the initialisation of the photometric calibration was
largely free of the effects of contamination by water ice \citep[see][for a summary of the
contamination problem in the early phases of the {\gaia} mission]{2016A&A...595A...1G}. As a
consequence the photometric system for \gdr{2} is different from \gdr{1}. This is illustrated in
\figref{fig:gdiffs} which shows the difference in $G$-band magnitude ($\Delta
G=G_\mathrm{DR1}-G_\mathrm{DR2}$) for the same sources between the two data releases. The source
pairs selected from the two releases match in celestial position to within $0.25$~arcsec and the
formal error on the magnitude differences is less than $0.01$. All sources were selected to have a
flux excess factor in \gdr{2} of less than $1.6$ (see \secref{sec:photdata} for a description of this
quantity). The two panels in \figref{fig:gdiffs} show that there is a substantial difference in the
$G$ band values, with the mean of $\Delta G$ being about $-0.1$~mag, and a strong colour dependence
which is indicated by the dashed line showing the polynomial relation
\begin{multline}
  G_\mathrm{DR1}-G_\mathrm{DR2} = 
  -0.013612 - 0.079627(\gbp-\grp) \\ - 0.0040444(\gbp-\grp)^2 +
  0.0018602(\gbp-\grp)^3\,.
  \label{eq:diffgvsbpminrp}
\end{multline}
Removing the colour dependence and plotting $\Delta G$ vs.\ $G$ (bottom panel of
\figref{fig:gdiffs}) reveals image saturation effects at the bright end which more strongly affect
the \gdr{1} magnitudes. Sources with larger magnitude differences typically have large estimated
uncertainties (blue points) whereas the majority of sources have smaller differences and small
estimated errors (red points). The feature near $G\approx11.5$ mag is due to the high and variable
photometric uncertainties in \gdr{1} for bright sources \citep[see figure 9 in][]{DR2-DPACP-40}.

This difference in photometric systems means that one should not apply photometric calibrations
derived from \gdr{1} \cite[e.g., the calibration of the red-clump absolute $G$-band
magnitude,][]{2018A&A...609A.116R, 2017MNRAS.471..722H} to \gdr{2} photometry.  The $G$ passband
calibration also changes from \gdr{1} to \gdr{2}. The passbands for $G$, {\gbp}, and {\grp} are
described in \cite{DR2-DPACP-40}. They are available in the version that was used for the \gdr{2}
data processing and in a revised version which was determined after the processing was finished (see
\secref{sec:photcaveats}). The revised passband should be used for precise photometric work based on
the fluxes listed in \gdr{2}. The nominal (pre-launch) passband as provided on the {\gaia} science
performance pages\footnote{https://www.cosmos.esa.int/web/gaia/science-performance} and independent
passband calibrations based on \gdr{1} \citep{2018arXiv180201667W, 2017A&A...608L...8M} should not
be used.  Likewise the nominal transformations between the {\gaia} broad-band photometry and other
photometric systems listed in \cite{2010A&A...523A..48J} should not be used. Refer to
\cite{DR2-DPACP-40} for the updated relations. To take full advantage of the high precision \gdr{2}
photometry, predictions of the {\gaia} broad-band magnitudes for stellar evolutionary tracks or
isochrones in the colour-magnitude diagram \citep[e.g.][]{2016ApJ...823..102C, 2017ApJ...835...77M}
should be updated.

\subsection{Source list evolution}
\label{sec:sourcelist}

The processing for a given data release starts with a task that groups individual {\gaia}
observations and links them to sources on the sky \citep[see][for a description of this
process]{DR2-DPACP-51,2016A&A...595A...3F}. The observations are linked to known sources, or sources are
newly `created' from the clustering of the observations around a celestial position where previously
no source was known to exist. This leads to a working catalogue of sources (hereafter called `the
source list') and their corresponding observations, which forms the basis for the subsequent data
processing. In this list the sources are assigned a {\gaia} source identifier which is intended to
be stable for every source. The algorithm that carries out the grouping and linking was much
improved at the beginning of the processing for \gdr{2}. The improved source list will lead to the
following changes in linking the observations to the source identifiers for a substantial fraction
of sources:
\begin{itemize}
  \item The merging of groups of observations previously linked to more than one source will lead to
    a new source associated to the merged observations (with a new source identifier) and the
    disappearance of the original sources (along with their source identifiers).
  \item The splitting of groups of observations previously linked to one source will lead to new
    sources associated to the split groups of observations (with new source identifiers) and the
    disappearance of the original source (along with its source identifier).
  \item The list of observations linked to a source may change (and hence the source characteristics
    may change), while the source identifier remains the same.
\end{itemize}
In the processing for \gdr{2} the number of changes of source identifiers (where the physical
source remains the same) is large. At magnitudes brighter than $G\approx16$ some 80--90 per cent of
the sources changed source identifier. At $G\approx18$~mag this reduces to some 20 per cent, going
down to zero source identifier changes around $G=20$~mag.

The consequence is that one should not blindly use the source identifier to look up sources from
\gdr{1} in \gdr{2}. Example applications we have in mind are the repeat of an analysis done with
the first data release using the new data and the retrieval of a list \gdr{1} sources,
cross-matched against some other survey, from the \gdr{2} tables. The recommendation is to treat
the source lists from the two releases as completely independent. An additional field will be added
to \gdr{2} and subsequent releases which specifies the {\gaia} {\em source name} as `{\gaia} DR$n$
\texttt{source\_id}'. The bare source identifier can be used for efficient queries of the large
{\gaia} data base, while the source name should always be specified (i.e., including the data
release number) when referring to the source in the literature. To facilitate the tracing of sources
from \gdr{1} to \gdr{2} a table is provided which lists for each \gdr{2} source the potential
matching sources in \gdr{1} (and vice versa). For the majority of sources (over $99$ per cent)
there is a one-to-one correspondence (although the source identifier can differ), but multiple
matches may occur and then it is up to the user of the {\gaia} data to make a judgement as to which
pair is the correct match (where the possible differences in the $G$-band magnitude should be kept
in mind).

The source list is expected to stabilise in future {\gaia} data releases with much less change
expected between \gdr{2} and {\gaia~DR3}. However some evolution of the source lists will take
place up to the final data release and we stress that a change in source {\em character} can always
occur as observations are added in future data releases (e.g., a stable source can turn into a
variable from one data release to the next).

%-------------------------------------------------------------------
%
% Using Gaia DR2 data: completeness and limitations
%
%-------------------------------------------------------------------

\section{Using \gdr{2} data: completeness and limitations}
\label{sec:gdrlimitations}

\gdr{2} represents a major advance compared to \gdr{1}, featuring new data types and a much
expanded and improved astrometric and photometric data set. Nevertheless this release is still
intermediate, based on only a limited amount ($\sim22$ months) of input data, and still suffers
from simplifications in the data processing that will introduce shortcomings in the calibrations
which in turn can introduce systematic errors. We summarise here the main limitations of
\gdr{2} which the user of the data should be aware of.

\subsection{\gdr{2} validation and source filtering}
\label{sec:gdrtwovalidation}

The validation of the \gdr{2} results followed the process described in
\cite{2016A&A...595A...2G}. We refer to the papers listed in \secsref{sec:dataprocessing} and
\ref{sec:gdrtwosummary} for full details on the validation of the data done at the level of the
individual data processing systems. The overall validation, assessing the combined results is
described in \cite{DR2-DPACP-39}. As was the case for \gdr{1} the results validation revealed no
problems that prevented a timely release of \gdr{2}, but filtering of the available data
processing outputs before their incorporation into \gdr{2} was still necessary. The level of
filtering is significantly reduced compared to that for \gdr{1} as can be appreciated from the
substantial increase in the number of sources for which astrometric and photometric data is
published. We summarise the filtering that was applied with the aim of providing a better
understanding of some of the survey characteristics.

\subsubsection{Astrometry}

For the astrometric data set the results were filtered by requiring that a source was observed by
{\gaia} at least five times (five focal plane transits), and that the astrometric excess noise and
the semi-major axis of the position uncertainty ellipse are less than 20 and 100~mas, respectively.
In addition within the astrometric solution pipeline the parallax and proper motions are determined
only for sources satisfying the requirement that they are brighter than $G=21$, that the number of
`visibility periods' used is at least 6 (a visibility period represents a group of observations
separated from other such groups by at least four days), and that the semi-major axis of the
5-dimensional uncertainty ellipse is below a magnitude dependent threshold.  We refer to
\cite{DR2-DPACP-51} for the details. For sources that do not meet these requirements only the
positions are reported in \gdr{2}.

\subsubsection{Photometry}

The photometric inputs were filtered as follows. Sources without a well-determined value for $G$ do
not appear in \gdr{2}. The photometry in the $G$, {\gbp}, or {\grp} bands is only reported if the
source was observed at least twice by {\gaia} in the respective bands. For the so-called 'bronze'
sources \citep[see \secref{sec:photdata} and][]{DR2-DPACP-44} no colour information (i.e. no {\gbp}
and {\grp}) is reported. This also holds for sources fainter than $G=21$~mag and sources for which
the flux excess factor is above 5. Hence \gdr{2} contains a substantial number of sources ($\sim300$
million) for which no colour information is available. Note however that the filtering on flux
excess factor was not applied to the variable source time series tables, hence there may be sources
that have no {\gbp} and/or {\grp} value listed but for which a light curve in {\gbp} and/or {\grp}
is nevertheless reported.

\subsubsection{Radial velocities}
\label{sec:vradfiltering}

For sources satisfying the following conditions no radial velocity is reported in \gdr{2}. The
source is fainter than $\grvs=14$ (the limit refers to the flux as actually measured in the RVS
band, not the provisional {\grvs} value mentioned in \secref{sec:rvsdata}); the fraction of transits
where the source was detected as having a double-lined spectrum was larger than $0.1$ (this removes
detected double-lined spectroscopic binaries); the uncertainty on the radial velocity is above
20~\kms; the effective temperature corresponding to the spectral template used to derive the radial
velocity is outside the range $3550$--$6900$~K. By construction the RVS data processing limited the
range of possible radial velocities to $|\vrad|<1000$~\kms. Special care was taken for the 613
sources that had measured radial velocities with absolute values above $500$~\kms. Because this
small subset can easily be contaminated by outliers caused by data processing limitations, their
spectra were visually inspected. Of these 613 sources, 202 were included in \gdr{2} as valid high
velocity sources, while the remainder were removed from the published catalogue. For sources with
radial velocities at absolute values below $500$~\kms\ visual inspection was not possible due to the
progressively (much) higher numbers. The users of \gdr{2} should thus be aware of the specific
selection applied to sources with $|\vrad|>500$~\kms. We refer to \cite{DR2-DPACP-54} for more
details on this issue.

\subsubsection{Variable stars}

During the variability analysis a strict internal filtering was applied to the quality of the
photometric time series (such as removing negative or unrealistically low flux values). This
was followed by a filtering of the classification results to reduce the contamination due to data
processing artefacts and confusion between variable types. The outputs from the specialised variable
star characterisation pipelines were filtered to remove sources for which the results of the light
curve analysis were not deemed reliable enough. This combination of filters reduced the number of
sources flagged as variable to the numbers listed in \tabref{tab:gdr2stats}. The reader interested
in using the variable star data set is strongly advised to consult \cite{DR2-DPACP-49} and
references therein, as well as the online documentation.

\subsubsection{Astrophysical parameters}

The astrophysical parameter results are only presented for sources brighter than $G=17$ (no fainter
sources were processed) and only for sources for which $G$, {\gbp}, and {\grp} are reported. Further
filtering was applied based on the quality of the various inputs to the astrophysical parameter
estimation, where particularly strict criteria were applied to the extinction and reddening
estimations. The details of the filtering applied to the astrophysical parameters are best
understood in conjunction with the description of how these parameters were estimated. Hence we
refer to \cite{DR2-DPACP-43} for the details (see also \secref{sec:apcaveats}).

\subsubsection{Solar system objects}

For the solar system data set the filtering on input data quality (internal to the processing
pipeline) was followed only by the removal of some SSO observations for which the relative flux
uncertainty in the $G$ band was larger than $0.1$. This mainly removes observations of the very
`fast' SSOs for which the observation window may be badly placed (causing flux loss) toward the end
of the focal plane transit. In addition a selection of the SSO observations was removed as well as
some individual sources \citep[see][for details]{DR2-DPACP-32}.

\subsubsection{Duplicated sources}

A global filter concerns the removal of duplicates of sources, which sometimes occur when the
observation to source matching process creates two clusters of detections that later turn out to
belong to the same source \citep[see][]{2016A&A...595A...2G, 2016A&A...595A...3F}. The
$47\,802\,437$ sources for which the duplicate was removed are indicated as such. The removal of
duplicates is done after the completion of the data processing. Hence the observations corresponding
to the removed component are effectively not used for, and do not appear in, the published
catalogue. In future \gaia\ data releases the duplicates are expected to be merged into a single
source.

\subsection{Survey completeness}

As can be appreciated from \figref{fig:gmaghistos} the completeness of the {\gaia} survey has much
improved for the second data release, being essentially complete between $G=12$ and $G=17$. The
completeness at the bright end has improved, although a fraction of the bright stars at $G<7$ is
still missing with no stars brighter than $G=1.7$~mag appearing in \gdr{2}.
\cite{2016A&A...595A...2G} extensively explain how the combination of the {\gaia} scan law coverage
of the sky over the period covered by \gdr{1} combined with the filtering applied to the
astrometric and photometric results leads to strips or holes with a lack of sources (see figures
11 and 12 in that paper). Although much reduced (as seen in \figref{fig:sourcedensity}), these
artefacts are still present in the \gdr{2} source list and start appearing at $G>17$.

We list here a number of more specific remarks on the completeness of \gdr{2}:
\begin{itemize}
  \item The completeness for high proper motion stars has significantly improved with respect to
    \gdr{1}, but it is estimated that some 17 per cent of high proper motion stars (with
    $\mu>0.6$~arcsec~yr$^{-1}$) are still missing (for various reasons).
  \item In crowded regions the capability to observe all stars is reduced
    \citep{2016A&A...595A...1G}. In combination with the still limited data treatment in crowded
    areas \citep[see section 6.2 in][]{2016A&A...595A...2G} this means that the survey limit in
    regions with densities above a few hundred thousand stars per square degree can be as bright as
    $G=18$.
  \item As described in \secref{sec:sciencedemos} the effective angular resolution of the \gdr{2}
    source list has improved to $\sim0.4$~arcsec, with incompleteness in close pairs of stars
    starting below about 2 arcsec.  Refer to \cite{DR2-DPACP-39} for details.
  \item We repeat that the radial velocity, astrophysical parameter and variable star data sets are
    far from complete with respect to the overall \gdr{2} catalogue (see
    \secref{sec:gdrtwosummary} above). In particular the radial velocities are only reported for a
    restricted range in effective temperatures (of the spectral templates, see
    \secref{sec:vradfiltering}) and the completeness of the radial velocity catalogue with
    respect to \gdr{2} varies from 60 to 80 per cent \citep{DR2-DPACP-54} over the range $G=4$ to
    $G=12$.
  \item The solar system object sample processed for \gdr{2} was pre-selected and is not a
    complete sample with respect to criteria like dynamics, type, category, etc. In addition bright
    SSOs ($G\lesssim10$) were removed from the published results because the astrometry in that
    brightness range is limited in quality by calibration uncertainties and systematics related to
    the apparent source size and motion on the sky (leading to the use of inadequate PSF models for
    the image centroiding).
\end{itemize}
For more detailed information on the completeness of \gdr{2} we refer to the individual data
processing papers and the overall validation paper \citep{DR2-DPACP-39}. No attempt was made at deriving
a detailed survey selection function.

\subsection{Limitations}

\subsubsection{Astrometry}
\label{sec:astrometry-limitations}

The astrometry in \gdr{2} represents a major improvement over \gdr{1} with an order of magnitude
improvement in the uncertainties at the bright end and a vast expansion of available parallaxes and
proper motions. In particular the individual uncertainties are much closer to having been drawn from
Gaussian distributions and the systematics in the parallax uncertainties are now generally below the
$0.1$~mas level \citep[as estimated from the analysis of QSO parallaxes,][]{DR2-DPACP-51}. However, the
users of the \gdr{2} astrometry should be aware of the following. There is an overall parallax
zeropoint of $\sim-0.03$~mas (as estimated from QSO parallaxes, in the sense of the \gdr{2}
parallaxes being too small) which the data have not been corrected for (see below), and the
astrometry shows systematics correlated to celestial position, source colour, and source magnitude.
Moreover the parallaxes and proper motions show significant spatial (i.e.\ source-to-source)
correlations of up to $0.04$~mas and $0.07$~{\masyr} over angular scales from $<1$ to $20$ degrees
\citep[see][for a more detailed characterisation of the spatial covariances]{DR2-DPACP-51}. These
regional systematics are visible in maps of average QSO parallaxes and in dense fields where the
large amount of sources allows to average the astrometric parameters and visualise the systematic
differences in, for example, the parallax zeropoint \citep{DR2-DPACP-51,DR2-DPACP-39}.

One might expect that the published parallax values would have been adjusted according to the global
zeropoint, however a deliberate choice was made not to apply any corrections to the \gdr{2}
astrometry. This is motivated by the fact that the value of the zeropoint depends on the sample
used to estimate its value \citep{DR2-DPACP-39}. The differences are related to the dependence of the
systematics in the astrometry on source position, colour, and magnitude, meaning that the zeropoint
for QSOs (faint, blue) may not be representative of the zeropoint for a sample of bright red stars.
In addition the correction of the global zeropoint would represent an arbitrary choice with respect
to the regional systematics which would be left uncorrected.

The astrometric uncertainties listed in \gdr{2} are derived from the formal uncertainties resulting
from the astrometric data treatment, and unlike for \gdr{1} these have not been externally
calibrated \citep[by comparison to the {\hip} data,][]{2016A&A...595A...4L}. At a late stage during
the preparation of \gdr{2} a bug was discovered in the astrometric processing software. This did not
significantly affect the astrometric parameters themselves but resulted in a serious underestimation
of the uncertainties for the bright sources ($G\lesssim 13$). Rather than recomputing the full
solution, with serious repercussions for the downstream processing and publication schedule, it was
decided to apply an approximate ad hoc correction to the uncertainties. The details of this are
described in appendix A of \cite{DR2-DPACP-51}. While the corrected (published) uncertainties are
thus approximately consistent with the residuals of the astrometric solution, comparisons with
external data show that they are still underestimated \citep{DR2-DPACP-39}. The underestimation is
moderate ($\sim7$--$10$\%) for faint sources ($G>16$) outside the Galactic plane, but may reach 30
to 50 per cent for sources of intermediate magnitude ($G\simeq 12$--$15$). At brighter magnitudes a
comparison with {\hip} data suggests that uncertainties are underestimated by no more than 25 per
cent \citep{DR2-DPACP-39}. No additional correction was made in the published data based on these
external comparisons, and users of the data may have to allow for it in their analyses.

The PSF model used in the pre-processing is essentially the same as that used for \gdr{1}, and the
iterative loop between the astrometric and photometric data treatment and the pre-processing is not
yet closed \citep[see section 6.1 and figure 10 in][]{2016A&A...595A...2G}. This implies that the
PSF calibrations and the subsequent determination of the source flux and location have not benefited
from better input astrometry and source colours. These inadequacies in the instrument calibration
have a particularly large impact on the astrometry of bright stars ($G\lesssim13$) which is visible
in the uncertainties being larger than those for somewhat fainter stars. In addition there may be a
systematic rotation of the proper motion system for the bright stars with respect to QSOs \citep[see
\tabref{tab:qualitystats} and][]{DR2-DPACP-51}, and the parallax zeropoint may be different.

\subsubsection{Photometry}
\label{sec:photcaveats}

The strongly varying photometric uncertainty at the bright end in $G$ and the bumps in the
uncertainty around $G\sim13$ and $G\sim16$ visible in the \gdr{1} data \citep{2016A&A...595A...2G,
2017A&A...600A..51E} are still present although in much reduced form \citep{DR2-DPACP-40}. The
uncertainties on {\gbp} and {\grp} as a function of magnitude are much smoother with the integrated
prism photometry being much less sensitive to instrument configuration changes.

The flux excess factor can take extreme values and it was decided not to publish colour information
for sources with a flux excess factor above 5 (this is a rather liberal filtering).  We recommend
that the value of the flux excess factor is used to clean samples of sources selected from \gdr{2}
from the most problematic cases, in particular if accurate colour information is important. The flux
excess factor has a dependence on $(\gbp-\grp)$, which any filtering should take into account. We
refer to \cite{DR2-DPACP-40}, \cite{DR2-DPACP-31}, and \cite{DR2-DPACP-51} for more detailed
recommendations on cleaning samples from the effects of the flux excess in the BP/RP bands.

Although not really a limitation in the photometric data, we nevertheless point out the following in
relation to the photometric zeropoints and passbands. The photometric zeropoints used to convert the
photometric fluxes into the magnitudes listed in \gdr{2} are derived from the passbands used
internal to the processing for this release. The calibration of these passbands was done in a
preliminary manner and they have been updated after the \gdr{2} processing was completed through
an analysis employing BP/RP spectra which were not available for the earlier calibrations.  The
magnitude zeropoints for the updated passbands differ by up to 3~mmag from those used to calculate
the \gdr{2} magnitudes \citep{DR2-DPACP-40}. As remarked in \secref{sec:dr2vsdr1}, for precision
photometric work the updated passbands should be used and then the difference in zeropoints should
be accounted for (by recalculating the magnitudes from the fluxes listed in \gdr{2}).

\subsubsection{Radial velocity data}

When using the radial velocities from \gdr{2} the following limitations should be taken into
account. Single-lined spectroscopic binaries have been treated as single stars and only the median
radial velocity, together with information on the scatter in the underlying (but unpublished) epoch
radial velocities, is provided. Double lined spectroscopic binaries which were detected as such were
not processed and are missing from the \gdr{2} radial velocity data set. Double lined spectroscopic
binaries with a weak secondary component are present in the catalogue and have also been treated as
single stars. No radial velocities have been determined for stars with detected emission lines and
there are no radial velocities for `cool' and `hot' stars (\secref{sec:vradfiltering}). Radial
velocities with absolute values above $500$~{\kms} should be treated with some care. Beyond this
limit clearly dubious values were filtered out of the catalogue but it is not guaranteed that all
remaining radial velocities above $+500$~{\kms} or below $-500$~{\kms} are reliable.

Through comparison with other radial velocity surveys it is concluded that the \gdr{2} radial
velocities are accurate to a few $100$~\ms, where systematic differences can be due to both \gdr{2}
and the other surveys. \cite{DR2-DPACP-54} show that while offsets are lower than $300$~\ms\ for
bright stars ($\grvs<10$),  a trend with magnitude is seen in all the comparisons with other
surveys, reaching $\sim500$~\ms\ at the faint end.

Finally, we note that \gdr{2} lists the atmospheric parameters (\teff, \logg, \feh) of the
spectral templates used in the derivation of the radial velocities through the cross-correlation
technique. Their values should {\em not} be used as estimates of the actual atmospheric
parameters of the stars, they are only provided as extra information to judge the quality of
the radial velocities.

\subsubsection{Astrophysical parameters}
\label{sec:apcaveats}

The values of \teff, \ag, \ebpminrp, radius, and luminosity were determined only from the three
broad-band photometric measurements and the parallax, on a star-by-star basis (where parallax was
not used to estimate {\teff}). The strong degeneracy between {\teff} and extinction/reddening when
using the broad band photometry necessitates rather extreme assumptions in order to estimate their
values. This can lead to correspondingly strong systematics in the astrophysical parameters which
are not accounted for in the uncertainties listed in \gdr{2}. We summarise here the most important
caveats but refer to the online documentation and \cite{DR2-DPACP-43} for more extensive guidelines
on the use of the astrophysical parameter estimates. The assessment of the quality of the
astrophysical parameters from the perspective of the overall validation of \gdr{2} can be found in
\cite{DR2-DPACP-39}.

The estimation of {\teff}, {\ag}, and {\ebpminrp} was done using a machine learning algorithm
\citep[specifically, the extremely randomised trees, or \textsc{ExtraTrees}
algorithm][]{Geurts2006}. For the {\teff} estimation the algorithm was trained on the photometry for
{\gaia} sources for which {\teff} estimates were available from existing independent surveys
\citep[see][table 2]{DR2-DPACP-43}. Only effective temperatures over the range $3000$--$10\,000$~K
were considered and the training data shows strong peaks at specific {\teff} values. The training
set for the extinction and reddening estimation was based on synthetic photometry constructed using
PARSEC 1.2S\footnote{\url{http://step.oapd.inaf.it/cgi-bin.cmd}} stellar models which are
accompanied by simulated photometry based on the Atlas 9 synthetic spectral library
\citep{2004astro.ph..5087C}. No attempt was made at a realistic population of the synthetic colour
magnitude diagrams in terms of the stellar initial mass function, the metallicity distribution, or
the frequency of extinction values. All sources were treated as single stars and no attempt was made
to filter out known galaxies, binaries, etc. Please refer to \cite{DR2-DPACP-43} for full details.

No {\teff} values outside the range $3000$--$10\,000$~K are reported as these were not contained in
the training data used for the estimation algorithm. This means that stars with effective
temperatures outside the aforementioned range will have systematically too high or too low {\teff}
values listed in \gdr{2}. The distribution of {\teff} values contains artefacts that reflect the
distribution of the {\teff} values in the training data. Effective temperature estimates in high
extinction areas can be underestimated as the training data contained no examples of extincted
stars.

The estimates of {\ag} and {\ebpminrp} have such large uncertainties in general that their
usefulness for individual stars is very limited. The extinction/reddening estimates should be used
statistically only (for collections of stars) in which case the extinction maps shown in
\cite{DR2-DPACP-43} demonstrate that on average the {\ag} estimates are reliable. The extinction
estimates are strictly non-negative (with a model grid imposed maximum of $\ag=4$) and have
non-Gaussian posteriors, for which asymmetric uncertainties are listed in the catalogue. The
non-negativity constraint can lead to apparent overestimation of the extinctions in regions, such as
at high Galactic latitudes, where low extinction is expected on average. The effective temperature
and extinction signals are degenerate in the broadband colours, which greatly limits the accuracy
with which either can be estimated.

The radius and luminosity are estimated from the value of {\teff} as determined from the {\gaia}
photometry, including a bolometric correction obtained from synthetic spectra. The resulting
estimates suffer from the naive use of $1/\varpi$ as a distance estimator and the assumption of zero
extinction. Their uncertainties are probably underestimated. 

\subsubsection{Variability data}

The variability data contained in \gdr{2} is somewhat complex and consists of three data sets, as
described in \secref{sec:varidata}, that overlap to a large degree \citep[for details refer
to][]{DR2-DPACP-49}. The mean $G$, {\gbp}, and {\grp} magnitudes and fluxes provided as part of the
light curve statistics can differ from the values provided in the overall \gdr{2} source table. In
these cases the median or mean magnitudes and fluxes from the variability data set are to be
preferred. There is a small number of stars with multiple entries in the SOS (Special Object
Studies) tables and there are sources with a different type in the SOS and automated variability
type estimation data sets.  Classifications different from those of independent variable star
surveys may occur \citep{DR2-DPACP-49, DR2-DPACP-39}.

%-------------------------------------------------------------------
%
% Using Gaia DR2 data: additional guidance
%
%-------------------------------------------------------------------
\section{Using \gdr{2} data: additional guidance}
\label{sec:guidance}

We briefly discuss a number of specific items that the users of \gdr{2} should keep in mind.
These concern issues inherent to the {\gaia} data (releases) and points to keep in mind when
interpreting the results from analyses of \gdr{2} data. More extensive examples of how to use the
data responsibly are provided in the papers listed at the start of \secref{sec:sciencedemos} and in
\cite{DR2-DPACP-38}.

\subsection{Time stamping in {\gaia} data releases}

\gdr{2} features photometric time series for sources varying in apparent magnitude and for solar
system objects, as well as astrometric time series for the latter. Future releases will in addition
contain time series for non-single star astrometry (such as binaries and stars with exoplanets),
radial velocities, and the medium and low resolution spectra from the RVS and BP/RP instruments. As
summarised in \cite{2016A&A...595A...4L} the primary coordinate system used for the {\gaia}
(astrometric) data processing is the Barycentric Celestial Reference System
\citep{2003AJ....126.2687S}. The BCRS is a relativistic reference system that is physically adequate
to describe both the motion of bodies in the solar system and the propagation of light from distant
celestial sources. The time-like coordinate of the BCRS is the barycentric coordinate time (TCB).
Consequently all the {\gaia} time series data are time-stamped using TCB. The numerical values in
the \gdr{2} tables are expressed JD$-2455197.5$(TCB) days, where by convention the origin for
{\gaia} time-stamping is J2010.0(TCB) = JD 2455197.5(TCB).

\subsection{Astrometric source model}

All sources were treated as single stars in the astrometric solution for \gdr{2}
\citep{DR2-DPACP-51}.  This means that physical binaries and multiple systems as well as extended
sources (galactic and extra-galactic, such as galaxies in the local universe), although present in
\gdr{2}, received no special treatment. Moreover the sources that are not single stars are not
marked as such. For binaries with orbital periods of the order of 2 years the proper motions or
parallaxes listed in \gdr{2} may be quite far from the true values for the system. The auxiliary
information in \gdr{2} can be used to isolate candidate non-single stars or galaxies but this should
be done with care and the results validated against known samples.

\subsection{Solar system object astrometry}

The epoch astrometry for SSOs is provided with uncertainties (on $\alpha, \delta$) and correlations.
These correlations are strong, reflecting the large difference in precision between the along-scan
and across-scan astrometric uncertainties which project into the uncertainties in $(\alpha,\delta)$
in a correlated manner. The correlations should be taken into account for any application in order
to recover the full accuracy of the astrometry in the along-scan direction. A known limitation of
asteroid astrometry in \gdr{2} is that the relativistic light deflection is computed as for the
stars (i.e., the source is considered to be at infinite distance). A correction corresponding to the
difference with respect to the finite distance must be applied whenever mas or sub-mas precision is
aimed at.

\subsection{Interpretation of photometric colours}

The problem of the excess flux in the BP/RP photometry manifests itself primarily at the faint
($G>19$) end of the survey, in crowded regions and around bright stars. In all these cases when
constructing colour magnitude diagrams one should be careful in interpreting them.

For example, open cluster sequences in non-crowded fields may manifest a turn towards the blue at
the lower end of the main sequence, which is a consequence of a stronger flux excess in BP than in
RP for faint sources. At the faint end one should be aware that the effects of zodiacal light are
clearly visible in the distribution of the flux excess factor \citep{DR2-DPACP-40}.

Care should be taken in the use of colour magnitude diagrams in crowded regions such as globular
cluster cores or the Milky Way bulge. Examples of colour-magnitude diagrams affected by the flux
excess problem are given in \cite{DR2-DPACP-39}. Finally, around bright sources there may be a
dependence in source colour on the distance from the bright source which will lead to spurious
features in a colour magnitude diagram.  

When faint red sources are being analysed it may be better to use the $(G-\grp)$ colour instead of
$(\gbp-\grp)$ as discussed in \cite{DR2-DPACP-31} for the case of brown dwarfs.

\subsection{Mean magnitudes of variable stars}

If a source is flagged as variable the recommendation is to use the mean value for its photometry
from the tables with variability information, as the varying brightness of the source can be more
carefully accounted for in the variability analysis.

\subsection{Use the astrophysical parameters with care}

\cite{DR2-DPACP-43} provide extensive guidance on the use of the astrophysical parameter estimates,
including how to select samples with the most reliable {\teff}, radius, and luminosity estimates,
and examples of how to use the estimates of {\ag} responsibly. We strongly recommend that these
guidelines are followed and encourage independent investigations into the quality and limitations of
the astrophysical parameter estimates.

\subsection{Filtering to create clean samples}

Although the bulk of the data in \gdr{2} is of excellent quality, specific analyses of the data
may require further filtering on data quality. One can find examples of how to do such filtering,
using the information contained in \gdr{2}, throughout the papers accompanying the release.
However, in many cases some experimentation by the user of the data will be needed to establish the
best ad-hoc filtering for a given application. Such filtering does come at the cost of introducing
additional truncation of the data which will further complicate the survey/sample selection function
and may in fact severely bias the interpretation of the results. For example, \cite{DR2-DPACP-33} show
how a seemingly innocuous selection on radial velocity error can lead to strong kinematic biases
when studying the Milky Way disk. Further examples of biases induced by sample truncation are given
in \cite{DR2-DPACP-38}. Finally, one should keep in mind that filtering on the observed values or
uncertainties of source parameters can increase the imprint on the resulting sample of, for example,
scanning law patterns.

\subsection{Negative parallaxes}

\gdr{2} represents the largest parallax catalogue ever produced and contains parallaxes of faint
objects observed relatively few times and of extragalactic objects. For many of such objects the
value of the parallax listed in the catalogue may be negative. As explained in \cite{DR2-DPACP-38} the
presence of negative parallaxes is a natural consequence of the way the {\gaia} observations are
described in terms of a linearised astrometric source model, with the parameters of the model solved
for through a least-squares process. Perhaps this is most easily appreciated by considering the
$0.5$ million QSOs appearing in \gdr{2} for which parallax solutions have been made. Given that
the true parallax for these sources is close to zero it is to be expected that for half of them the
observed parallax (as solved for from the observations) is negative (where in the case of \gdr{2}
the fraction of negative parallaxes for QSOs is higher because of the negative parallax zeropoint). 

Hence negative parallaxes represent perfectly valid measurements and can be included in analyses of
the \gdr{2} data. Examples of how one can do this are given in \cite{DR2-DPACP-38}.

\subsection{Known spurious results}

There are a number of results listed in \gdr{2} which are obviously wrong and which may surprise
the user of the data. We point out two specific cases here.

For a small number of sources the parallaxes listed in \gdr{2} have very large positive or
negative values (with for example 59 sources having parallaxes larger than that of Proxima
Centauri), where the negative values can be very far from zero when expressed in terms of the formal
uncertainty on the parallax. These parallax values are spurious and caused by a close alignment (of
order $0.2$--$0.3$ arcsec) of sources, that are only occasionally resolved in the {\gaia}
observations, depending on the scan direction. These cases show up typically in dense regions
covered by only a few transits or an unfortunate distribution of scan directions and parallax
factors. This is consistent with most of these sources being faint and concentrated in dense areas
along the Galactic plane and toward the Galactic bulge \citep[see figure C4 in][]{DR2-DPACP-51}. Most
likely the proper motions of these sources are also erroneous. This is consistent with the presence
of a number of high-proper motion stars at $G>19$ ($104\,243$ at $\mu>100$~\masyr, $12\,431$ at
$\mu>200$~\masyr, and $4459$ at $\mu>300$~\masyr) which show a marked concentration toward the
galactic bulge and galactic plane regions. These sources overlap to a large degree with the sources
with spurious parallax values and their proper motions are thus likely to be unreliable. More
details on this problem and guidance on how to clean samples from spurious parallaxes can be found
in \cite{DR2-DPACP-51} (in particular their appendix C).

Among the bright and well known (i.e.\ named) variable stars there are a number of cases where the
mean $G$-band magnitude listed in \gdr{2} is clearly wrong. One prominent case is the star RR
Lyrae itself for which the mean magnitude is listed as $G=17$. The wrong value is caused by the fact
that the treatment of outliers, as implemented in the photometric processing for \gdr{2}, is not
efficient in the case of variable sources that have an intrinsically large spread in the individual
photometric observations. As a consequence of the wrong magnitude estimate, the parallax of RR Lyrae
was determined to be $-2.6$ mas.

We stress here that the above problems concern only a very small number of cases which do not
indicate overall problems with the quality of \gdr{2}.

\subsection{Take into account uncertainties and correlations}

The astrometric uncertainties are provided in the form of the full covariance matrix for the five
astrometric parameters. The correlations between the uncertainties can be significant and they
should always be accounted for to correctly calculate the standard uncertainties on linear
combinations of (subsets of) the astrometric parameters and to correctly assess, for example, how
far away a given set of astrometric parameters is from a model prediction. The mathematics involved
in accounting for correlated uncertainties is summarised in \cite{DR2-DPACP-38} and described more
extensively in the \gdr{2} online documentation.

In this context we point out that the longest principal axis of a scaled version of the covariance
matrix is provided as the parameter \texttt{astrometric\_sigma\_5dmax} for both the 5-parameter and
2-parameter solutions. This parameter is equivalent to the semi-major axis of the position error
ellipse and can be useful in filtering out sources for which one of the astrometric parameters, or a
linear combination of several parameters, is particularly ill-determined. We refer to the online
\gdr{2} documentation for more details.

\subsection{Dealing with underestimated uncertainties and/or systematic errors}

As pointed out above the uncertainties quoted in \gdr{2} on the various source parameters can be
underestimated and there are also systematic errors with varying dependencies on source brightness,
colour, and position on the sky, which moreover may be spatially correlated. We can provide no
general recipe for taking these effects into account in scientific analyses of the \gdr{2} data,
but give a few recommendations here.

We strongly advise against attempts to `correct' the data themselves as a means to get rid of
underestimated uncertainties or systematic errors. This would require a level of understanding and
characterisation of these effects that would have allowed their removal during the data processing
in the first place. We recommend (for studies where it matters) to include the presence of
systematic effects in the uncertainties as part of the data analysis, for example in a forward
modelling approach. The level of systematic errors (e.g.\ the size of the parallax zeropoint) or the
factors by which uncertainties are under- or overestimated then become part of the model parameters
to estimate.  Examples of such analyses of \gdr{1} parallax data can be found in
\cite{2017A&A...599A..67C} and \cite{2017ApJ...838..107S}, where the latter include both a parallax
zeropoint and a scaling factor for the quoted uncertainties as part of their probabilistic model
that fits a period luminosity relation to data for RR Lyrae stars. The spatial correlation
parameters for the uncertainties and systematic errors can be included in a similar way as part of
the modelling. Further guidance on the use of the astrometric data (in particular the parallaxes)
from \gdr{2} can be found in \cite{DR2-DPACP-38}.

%-------------------------------------------------------------------
%
% Gaia DR2 access facilities
%
%-------------------------------------------------------------------

\section{\gdr{2} access facilities}
\label{sec:access}

The main entry point to \gdr{2} remains the ESA {\gaia} archive, which can be accessed at
\url{http://archives.esac.esa.int/gaia}. Access is also possible through a number of partner and
affiliate data centres in Europe, the United States, Japan, Australia, and South Africa. These data
centres provide their own access facilities, but do not necessarily host all data contained in the
ESA {\gaia} archive. The services offered at the ESA {\gaia} archive remain as described in
\cite{2016A&A...595A...2G} and we list here a few enhancements and changes.
\begin{itemize}
  \item The access to the light curves for variable stars is now in the form of a URL that links
    from the main \texttt{gaia\_source} table to the specific files that contain the light curves
    for the source in VOTable format\footnote{\url{http://www.ivoa.net/documents/latest/VOT.html}}.
  \item The astrometric and photometric time series for the SSOs are all collated into one large
    table containing multiple entries for each SSO. Note that the source identifiers for SSOs are
    negative numbers. To enable queries of SSOs based on orbital elements or absolute magnitude, an
    auxiliary table containing such data, plus ancillary quantities, is provided. In addition a
    table with the residuals of each Gaia observation with respect to an orbital fit is provided
    as a reference.
  \item The archive visualisation service \citep{2017A&A...605A..52M} has been much expanded to
    allow for efficient preliminary exploration of the data in the entire \gdr{2} catalogue. The
    service offers several pre-computed diagrams which can be explored through linked views and
    allows one to interactively define a query for a given data set. This serves in particular to
    narrow down queries for data to the exact samples one is interested in and thus save time and
    storage space for the actual query. Full details can be found in \cite{DR2-DPACP-42}.
  \item We provide pre-computed cross-matches between \gdr{2} and a number of other large surveys.
    We recommend using these cross-matches as they have been carefully validated and their use
    facilitates reproducing analyses of \gdr{2} data combined with other survey data. The details
    are provided in \cite{DR2-DPACP-41}. The pre-computed cross-matches are provided for the
    following surveys: {\hip} \citep[new reduction,][]{book:newhip}; {\tyc}-2
    \citep{2000A&A...355L..27H}; 2MASS \citep{2006AJ....131.1163S}; SDSS DR9
    \citep{2012ApJS..203...21A}; APASS DR9 \citep{apass9, 2015AAS...22533616H}; UCAC4
    \citep{2013AJ....145...44Z}; Pan-STARRS1 \citep{2016arXiv161205560C}; AllWise
    \citep{2010AJ....140.1868W}; GSC2.3 \citep{2008AJ....136..735L}; PPMXL
    \citep{2010AJ....139.2440R}; URAT1 \citep{2015AJ....150..101Z}; and RAVE DR5
    \citep{2017AJ....153...75K}.
\end{itemize}
Finally we mention the creation of a {\gaia} Community forum
(\url{https://www.cosmos.esa.int/web/gaia/forum}) which is intended to facilitate discussion on the
use of {\gaia} data. The principle is to let the users of the data discuss amongst themselves on
this forum but the discussions will be monitored by members from the {\gaia} Data Processing and
Analysis Consortium who may respond with comments and expert advice when necessary.

%-------------------------------------------------------------------
%
% Conclusions
%
%-------------------------------------------------------------------

\begin{figure}[t]
  \resizebox{\hsize}{!}{\includegraphics{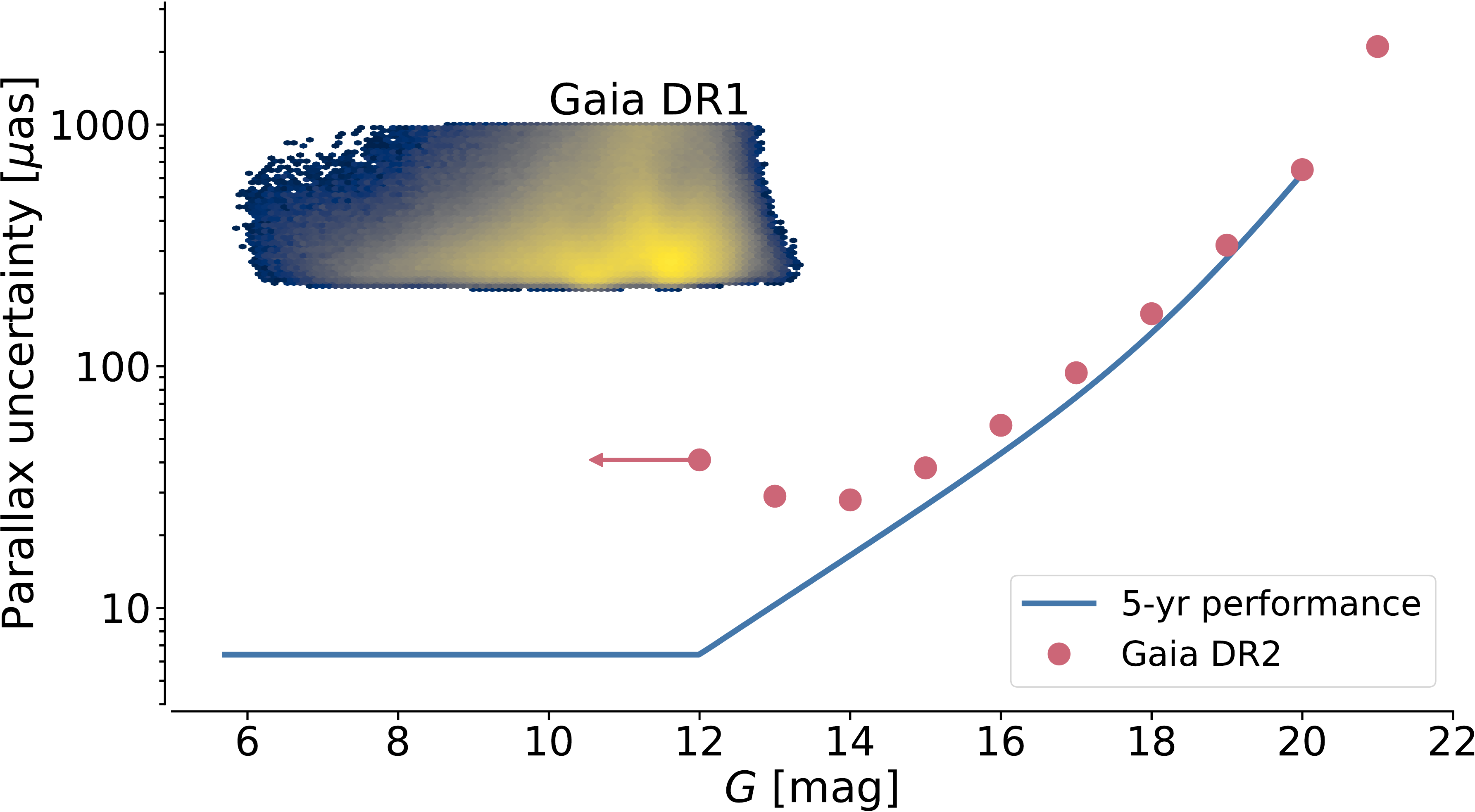}}
  \caption{Parallax uncertainties in \gdr{2} (dots) as a function of $G$ compared to the
  uncertainties quoted for \gdr{1} (colour scale) and the expected end-of-mission parallax
  performance (solid line), as predicted after the commissioning of {\gaia}. Note how the
  performance for \gdr{2} is still limited by calibration uncertainties for sources brighter than
  $G\sim14$.\label{fig:dr2dr1sigvarpi}}
\end{figure}

\section{Conclusions}
\label{sec:conclusions}

With the first {\gaia} data release in 2016 the astronomical community got an early taste of the
potential of the {\gaia} mission results, in particular through the 2 million parallaxes and proper
motions made available as part of the {\tyc}-{\gaia} Astrometric Solution. The science done with
\gdr{1} spans a wide range of topics and often features the powerful combination of {\gaia} and
other surveys. \gdr{1} was also quickly established as a reference for the astrometric and
photometric calibration of other surveys, resulting among others in the rejuvenation of existing
proper motion catalogues. For a brief review of the impact of \gdr{1} we refer to
\cite{2017arXiv170901216B}.

With the release of \gdr{2} the promise of the availability of fundamental astrophysical
information for (over) a billion sources spread over a substantial fraction of the volume of the
Milky Way starts to be fulfilled. The addition of the largest radial velocity survey to date,
coupled with astrophysical information for $161$ million sources and variability information for
half a million sources will make \gdr{2} a resource to be mined for stellar physics and galactic
as well as extra-galactic astronomy for many years to come. Moreover, \gdr{2} provides a first
glimpse of the immense power of {\gaia} for solar system studies.

Nevertheless \gdr{2} still represents an early data release based on only a limited amount (less
than two years) of input data, partly inadequate calibrations, and an incomplete understanding of
the behaviour of the spacecraft, payload, and instruments. These shortcomings manifest themselves as
systematic errors which although much reduced in size from \gdr{1} to \gdr{2} will remain a limiting
factor in scientific uses of the data, in particular at the bright end of the survey and, for
example, for distant samples. This is illustrated in \figref{fig:dr2dr1sigvarpi} which shows the
parallax uncertainties as a function of $G$ for \gdr{2} (dots), \gdr{1} (colour scale map), and the
end of mission \citep[solid line, as predicted after {\gaia} commissioning,][]{2016A&A...595A...1G}.
The bright end ($G\lesssim14$) performance for \gdr{2} is still limited by calibration errors, while
at the faint end the nominal end of mission performance is already being reached (this is probably
due to a conservative assessment of the effect at the faint end of the excess stray light). The task
for the {\gaia} data processing for the next data release will thus be to substantially reduce the
systematics such that a real advantage can be gained, in particular at the bright end, from the
increase in precision due to the longer time span of the input data. The main challenges will be the
following. The PSF modelling used in the image location determination must be upgraded, such that
for example astrometric colour terms are already accounted for at an early stage. This also requires
the closing of the iterative loop shown in figure 10 in \cite{2016A&A...595A...2G}. The modelling of
the sky background (both astronomical and as caused by the excess stray light on board {\gaia}) has
to be refined to further improve the image location process and to get rid of the flux excess in the
BP/RP photometry. The latter will also benefit from an improvement in the treatment of crowded
fields, specifically a better treatment of the effects of overlapping images in all of {\gaia}'s
instruments and in particular for the BP/RP/RVS instruments where the measurement of spectra
necessitates much larger images in the focal plane.  Finally the origins of the systematic effects
in the astrometry will be further investigated with much effort to be dedicated to the continued
development of the possibility to calibrate the systematic effects from the observations.

The next {\gaia} data release will also feature new data products of which the BP/RP and RVS spectra
and the non-single star astrometric and radial velocity solutions represent qualitative changes in
the character with respect to \gdr{2}. Further enhancements include: epoch astrometry for
non-single stars, an expanded radial velocity survey (to $\grvs\sim14$) including the analysis of
spectroscopic binaries, astrophysical parameter estimates based on BP/RP/RVS spectra, a further
order of magnitude increase in the availability of variability information, the first results from
eclipsing binary star processing, analyses of extended objects (galaxies, QSO hosts), and an expanded
list of some hundred thousand solar system objects for which multi-colour photometry will also be
provided. The latter opens up for investigation the powerful combination of precise orbits for
SSOs combined with a homogeneous multi-colour photometric survey of these bodies.

Hence there is much more to come from {\gaia} but for now we invite the reader to start exploring
the magnificent survey that is \gdr{2}.

%-------------------------------------------------------------------
%
% Acknowledgements
%
%-------------------------------------------------------------------

\begin{acknowledgements}

This work presents results from the European Space Agency (ESA) space mission \gaia. \gaia\ data are
being processed by the \gaia\ Data Processing and Analysis Consortium (DPAC). Funding for the DPAC
is provided by national institutions, in particular the institutions participating in the \gaia\
MultiLateral Agreement (MLA). The \gaia\ mission website is \url{https://www.cosmos.esa.int/gaia}.
The \gaia\ Archive website is \url{http://gea.esac.esa.int/archive/}.

The \gaia\ mission and data processing have financially been supported by, in alphabetical order by country:
the Algerian Centre de Recherche en Astronomie, Astrophysique et G\'{e}ophysique of Bouzareah Observatory;
the Austrian Fonds zur F\"{o}rderung der wissenschaftlichen Forschung (FWF) Hertha Firnberg Programme through grants T359, P20046, and P23737;
the BELgian federal Science Policy Office (BELSPO) through various PROgramme de D\'eveloppement
d'Exp\'eriences scientifiques (PRODEX) grants and the Polish Academy of Sciences - Fonds
Wetenschappelijk Onderzoek through grant VS.091.16N;
the Brazil-France exchange programmes Funda\c{c}\~{a}o de Amparo \`{a} Pesquisa do Estado de S\~{a}o
Paulo (FAPESP) and Coordena\c{c}\~{a}o de Aperfeicoamento de Pessoal de N\'{\i}vel Superior (CAPES)
- Comit\'{e} Fran\c{c}ais d'Evaluation de la Coop\'{e}ration Universitaire et Scientifique avec le
Br\'{e}sil (COFECUB);
the Chilean Direcci\'{o}n de Gesti\'{o}n de la Investigaci\'{o}n (DGI) at the University of Antofagasta and the Comit\'e Mixto ESO-Chile;
the National Science Foundation of China (NSFC) through grants 11573054 and 11703065;  
the Czech-Republic Ministry of Education, Youth, and Sports through grant LG 15010, the Czech Space
Office through ESA PECS contract 98058, and Charles University Prague through grant PRIMUS/SCI/17;    
the Danish Ministry of Science;
the Estonian Ministry of Education and Research through grant IUT40-1;
the European Commission’s Sixth Framework Programme through the European Leadership in Space
Astrometry (\href{https://www.cosmos.esa.int/web/gaia/elsa-rtn-programme}{ELSA}) Marie Curie
Research Training Network (MRTN-CT-2006-033481), through Marie Curie project PIOF-GA-2009-255267
(Space AsteroSeismology \& RR Lyrae stars, SAS-RRL), and through a Marie Curie Transfer-of-Knowledge
(ToK) fellowship (MTKD-CT-2004-014188); the European Commission's Seventh Framework Programme
through grant FP7-606740 (FP7-SPACE-2013-1) for the \gaia\ European Network for Improved data User
Services (\href{https://gaia.ub.edu/twiki/do/view/GENIUS/WebHome}{GENIUS}) and through grant 264895 for the \gaia\
Research for European Astronomy Training
(\href{https://www.cosmos.esa.int/web/gaia/great-programme}{GREAT-ITN}) network;
the European Research Council (ERC) through grants 320360 and 647208 and through the European
Union’s Horizon 2020 research and innovation programme through grants 670519 (Mixing and Angular
Momentum tranSport of massIvE stars -- MAMSIE) and 687378 (Small Bodies: Near and Far);
the European Science Foundation (ESF), in the framework of the \gaia\ Research for European
Astronomy Training Research Network Programme
(\href{https://www.cosmos.esa.int/web/gaia/great-programme}{GREAT-ESF});
the European Space Agency (ESA) in the framework of the \gaia\ project, through the Plan for
European Cooperating States (PECS) programme through grants for Slovenia, through contracts C98090
and 4000106398/12/NL/KML for Hungary, and through contract 4000115263/15/NL/IB for Germany;
the European Union (EU) through a European Regional Development Fund (ERDF) for Galicia, Spain;    
the Academy of Finland and the Magnus Ehrnrooth Foundation;
the French Centre National de la Recherche Scientifique (CNRS) through action 'D\'efi MASTODONS',
the Centre National d'Etudes Spatiales (CNES), the L'Agence Nationale de la Recherche (ANR)
'Investissements d'avenir' Initiatives D’EXcellence (IDEX) programme Paris Sciences et Lettres
(PSL$\ast$) through grant ANR-10-IDEX-0001-02, the ANR 'D\'{e}fi de tous les savoirs' (DS10)
programme through grant ANR-15-CE31-0007 for project 'Modelling the Milky Way in the Gaia era'
(MOD4Gaia), the R\'egion Aquitaine, the Universit\'e de Bordeaux, and the Utinam Institute of the
Universit\'e de Franche-Comt\'e, supported by the R\'egion de Franche-Comt\'e and the Institut des
Sciences de l'Univers (INSU);
the German Aerospace Agency (Deutsches Zentrum f\"{u}r Luft- und Raumfahrt e.V., DLR) through grants
50QG0501, 50QG0601, 50QG0602, 50QG0701, 50QG0901, 50QG1001, 50QG1101, 50QG1401, 50QG1402, 50QG1403,
and 50QG1404 and the Centre for Information Services and High Performance Computing (ZIH) at the
Technische Universit\"{a}t (TU) Dresden for generous allocations of computer time;
the Hungarian Academy of Sciences through the Lend\"ulet Programme LP2014-17 and the J\'anos Bolyai
Research Scholarship (L.~Moln\'ar and E.~Plachy) and the Hungarian National Research, Development,
and Innovation Office through grants NKFIH K-115709, PD-116175, and PD-121203;
the Science Foundation Ireland (SFI) through a Royal Society - SFI University Research Fellowship (M.~Fraser);
the Israel Science Foundation (ISF) through grant 848/16;
the Agenzia Spaziale Italiana (ASI) through contracts I/037/08/0, I/058/10/0, 2014-025-R.0, and
2014-025-R.1.2015 to the Italian Istituto Nazionale di Astrofisica (INAF), contract 2014-049-R.0/1/2
to INAF dedicated to the Space Science Data Centre (SSDC, formerly known as the ASI Sciece Data
Centre, ASDC), and contracts I/008/10/0, 2013/030/I.0, 2013-030-I.0.1-2015, and 2016-17-I.0 to the
Aerospace Logistics Technology Engineering Company (ALTEC S.p.A.), and INAF;
the Netherlands Organisation for Scientific Research (NWO) through grant NWO-M-614.061.414 and
through a VICI grant (A.~Helmi) and the Netherlands Research School for Astronomy (NOVA);
the Polish National Science Centre through HARMONIA grant 2015/18/M/ST9/00544 and ETIUDA grants 2016/20/S/ST9/00162 and 2016/20/T/ST9/00170;
the Portugese Funda\c{c}\~ao para a Ci\^{e}ncia e a Tecnologia (FCT) through grant
SFRH/BPD/74697/2010; the Strategic Programmes UID/FIS/00099/2013 for CENTRA and UID/EEA/00066/2013
for UNINOVA;
the Slovenian Research Agency through grant P1-0188;
the Spanish Ministry of Economy (MINECO/FEDER, UE) through grants ESP2014-55996-C2-1-R,
ESP2014-55996-C2-2-R, ESP2016-80079-C2-1-R, and ESP2016-80079-C2-2-R, the Spanish Ministerio de
Econom\'{\i}a, Industria y Competitividad through grant AyA2014-55216, the Spanish Ministerio de
Educaci\'{o}n, Cultura y Deporte (MECD) through grant FPU16/03827, the Institute of Cosmos Sciences
University of Barcelona (ICCUB, Unidad de Excelencia 'Mar\'{\i}a de Maeztu') through grant
MDM-2014-0369, the Xunta de Galicia and the Centros Singulares de Investigaci\'{o}n de Galicia for
the period 2016-2019 through the Centro de Investigaci\'{o}n en Tecnolog\'{\i}as de la
Informaci\'{o}n y las Comunicaciones (CITIC), the Red Espa\~{n}ola de Supercomputaci\'{o}n (RES)
computer resources at MareNostrum, and the Barcelona Supercomputing Centre - Centro Nacional de
Supercomputaci\'{o}n (BSC-CNS) through activities AECT-2016-1-0006, AECT-2016-2-0013,
AECT-2016-3-0011, and AECT-2017-1-0020;
the Swedish National Space Board (SNSB/Rymdstyrelsen);
the Swiss State Secretariat for Education, Research, and Innovation through the ESA PRODEX
programme, the Mesures d’Accompagnement, the Swiss Activit\'es Nationales Compl\'ementaires, and the
Swiss National Science Foundation;
the United Kingdom Rutherford Appleton Laboratory, the United Kingdom Science and Technology
Facilities Council (STFC) through grant ST/L006553/1, the United Kingdom Space Agency (UKSA) through
grant ST/N000641/1 and ST/N001117/1, as well as a Particle Physics and Astronomy Research Council
Grant PP/C503703/1.

The GBOT programme \citep{2016A&A...595A...1G, 2014SPIE.9149E..0PA} uses observations collected at
(i) the European Organisation for Astronomical Research in the Southern Hemisphere (ESO) with the
VLT Survey Telescope (VST), under ESO programmes 092.B-0165, 093.B-0236, 094.B-0181, 095.B-0046,
096.B-0162, 097.B-0304, 098.B-0034, 099.B-0030, 0100.B-0131, and 0101.B-0156, and (ii) the Liverpool
Telescope, which is operated on the island of La Palma by Liverpool John Moores University in the
Spanish Observatorio del Roque de los Muchachos of the Instituto de Astrof\'{\i}sica de Canarias
with financial support from the United Kingdom Science and Technology Facilities Council, and (iii)
telescopes of the Las Cumbres Observatory Global Telescope Network.

In this work we made use of the Set of Identifications, Measurements, and Bibliography for
Astronomical Data \citep[SIMBAD;][]{2000A&AS..143....9W}, the 'Aladin sky atlas'
\citep{2000A&AS..143...33B,2014ASPC..485..277B}, and the VizieR catalogue access tool
\citep{2000A&AS..143...23O}, all operated at the Centre de Donn\'ees astronomiques de Strasbourg
(\href{http://cds.u-strasbg.fr/}{CDS}). We additionally made use of Astropy, a community-developed
core Python package for Astronomy \citep{2018arXiv180102634T}, IPython \citep{PER-GRA:2007},
Matplotlib \citep{2007CSE.....9...90H}, and TOPCAT
\citep[\url{http://www.starlink.ac.uk/topcat/}]{2005ASPC..347...29T}.

We thank the anonymous referee for suggestions that greatly helped improve the readability and
clarity of this paper.
\end{acknowledgements}

%-------------------------------------------------------------------
%
% Bibliography
%
%-------------------------------------------------------------------

\bibliographystyle{aa} % style aa.bst
\bibliography{refsdrtwo} % your references refs.bib

%-------------------------------------------------------------------
%
% Appendices
%
%-------------------------------------------------------------------

\begin{appendix}

  \section{List of acronyms}

  \begin{table}[h]
    \caption{List of acronyms used in this paper.\label{app:acronyms}}
    \begin{tabular}{ll}
      \hline\hline
      \noalign{\smallskip}
      Acronym & Description \\
      \noalign{\smallskip}
      \hline
      \noalign{\smallskip}
      2MASS&Two-Micron All Sky Survey \\
      AAVSO&American Association of Variable Star Observers \\
      APASS&AAVSO Photometric All-Sky Survey \\
      BCRS&Barycentric Celestial Reference System \\
      BP&Blue Photometer \\
      CCD&Charge-Coupled Device \\
      DPAC&Data Processing and Analysis Consortium \\
      ESA&European Space Agency \\
      GBOT&Ground-Based Optical Tracking \\
      GSC&Guide Star Catalog \\
      ICRF&International Celestial Reference Frame \\
      JD&Julian Date \\
      LMC&Large Magellanic Cloud \\
      OBMT&On-Board Mission Timeline \\
      PSF&Point Spread Function \\
      PPMXL&Position and Proper Motion Extended-L Catalog \\
      QSO&Quasi-Stellar Object \\
      RMS&Root-Mean-Square \\
      RP&Red Photometer \\
      RVS&Radial Velocity Spectrometer \\
      SDSS&Sloan Digital Sky Survey \\
      SMC&Small Magellanic Cloud (special, high-density area on the sky) \\
      SOS&Specific Object Studies \\
      SSO&Solar-System Object \\
      TCB&Barycentric Coordinate Time \\
      TGAS&Tycho-Gaia Astrometric Solution \\
      UCAC&USNO CCD Astrograph Catalog \\
      URAT&USNO Robotic Astrometric Telescope \\
      URL&Uniform Resource Locator \\
      \noalign{\smallskip}
      \hline
    \end{tabular}
  \end{table}

\end{appendix}

\end{document}